\documentclass[sigconf, nonacm, pdfa]{acmart}
\usepackage[a-2b]{pdfx}

\usepackage{enumerate}
\usepackage{listings}
\usepackage[linesnumbered, ruled, vlined,resetcount]{algorithm2e} % For algorithms
\usepackage{graphics}
\usepackage{balance}
\usepackage{graphicx}
\usepackage{epstopdf}
\usepackage{subfig}
\usepackage{amsthm}
\usepackage{bm}
\usepackage{algpseudocode}
\usepackage{listings}
\usepackage[ruled, vlined, linesnumbered]{algorithm2e}
\usepackage{wrapfig}
\usepackage{tikz}
\usepackage{pifont}
\usepackage{enumitem}
\usepackage{booktabs}
\usepackage{subfig}
\usepackage[ruled, vlined, linesnumbered, resetcount]{algorithm2e}
\usepackage{hyperref}
\usepackage{siunitx}
\usepackage{multirow}
\usepackage{makecell}

\newcommand\vldbdoi{10.14778/3704965.3704986}
\newcommand\vldbpages{4827 - 4839}
\newcommand\vldbvolume{17}
\newcommand\vldbissue{13}
\newcommand\vldbyear{2024}
\newcommand\vldbauthors{\authors}
\newcommand\vldbtitle{\shorttitle} 
\newcommand\vldbavailabilityurl{URL_TO_YOUR_ARTIFACTS}
\newcommand\vldbpagestyle{empty}

\newcommand{\lhf}[1]{\textcolor{blue}{#1}}

\newcommand{\ys}[1]{\textcolor{blue}{#1}}
\newcommand{\todo}[1]{\textcolor{green}{#1}}

\newcommand{\ie}{\emph{i.e.,}\xspace}
\newcommand{\eg}{\emph{e.g.,}\xspace}

\newcommand{\oursys}{\texttt{GastCoCo}\xspace}
\newcommand{\ourstorage}{\texttt{CBList}\xspace}
\newcommand{\PrAllNeis}{\textit{scan\_edges($v_{src}$)}\xspace}
\newcommand{\PrOneNei}{\textit{read\_edge($v_{src}$,$v_{dst}$)}\xspace}
\newcommand{\PrAllVs}{\textit{scan\_vertices()}\xspace}
\newcommand{\PrAllVsCond}{\textit{scan\_vertices(cond)}\xspace}
\newcommand{\PrOneV}{\textit{read\_vertex(v)}\xspace}

\newcommand{\etitle}[1]{\vspace{1.2ex}\noindent{\em\underline{#1}}}
\newcommand{\eat}[1]{}

\usepackage{tikz}
\newcommand{\blackcircleone}{%
\tikz[baseline=(char.base)]{
    \node[shape=circle,fill=black,draw,text=white,inner sep=0.5pt] (char) {1};
}}
\newcommand{\blackcircletwo}{%
\tikz[baseline=(char.base)]{
    \node[shape=circle,fill=black,draw,text=white,inner sep=0.5pt] (char) {2};
}}
\newcommand{\blackcirclethree}{%
\tikz[baseline=(char.base)]{
    \node[shape=circle,fill=black,draw,text=white,inner sep=0.5pt] (char) {3};
}}
\newcommand{\blackcirclefour}{%
\tikz[baseline=(char.base)]{
    \node[shape=circle,fill=black,draw,text=white,inner sep=0.5pt] (char) {4};
}}

\newcommand{\GVc}{\texttt{GetVertices($cond$)}\xspace}

\newcommand{\GEv}{\texttt{GetNeighbors($vertex$)}\xspace}
\newcommand{\GEc}{\texttt{GetNeighbors($chain$)}\xspace}
\newcommand{\FE}{\texttt{FindNeighbor($edge$)}\xspace}

\begin{document}
% \title{CBList: A graph dynamic storage that supports graph analysis}
% \title{SoCo: Graph Storage and Software Prefetch Co-Design for Dynamic Graph Computation}
\title{GastCoCo:
  Graph Storage and Coroutine-Based Prefetch Co-Design for Dynamic Graph Processing}
% \title{GSCO$^2$: Graph Storage and Coroutine-Based Prefetch Co-Design for Dynamic Graph Processing}
%\title{A Prefetch-Friendly Graph Index Design for Dynamic Graph Processing}
%%
%% The "author" command and its associated commands are used to define the authors and their affiliations.
\author{Hongfu Li$^*$}

\affiliation{%
  \institution{Northeastern Univ., China} 
  % \streetaddress{}
  % \city{Shenyang}
  % \state{China}
  % \postcode{110000}
}
\email{lihongfu@stumail.neu.edu.cn}

\author{Qian Tao$^*$}
\affiliation{%
  \institution{Tongyi Lab, Alibaba Group}
}
\email{qian.tao@alibaba-inc.com}

\author{Song Yu}
\affiliation{%
  \institution{Northeastern Univ., China} 
  % \streetaddress{}
  % \city{Shenyang}
  % \state{China}
  % \postcode{110000}
}
\email{yusong@stumail.neu.edu.cn}

\author{Shufeng Gong}
\affiliation{%
  \institution{Northeastern Univ., China} 
  % \streetaddress{}
  % \city{Shenyang}
  % \state{China}
  % \postcode{110000}
}
\email{gongsf@mail.neu.edu.cn}

\author{Yanfeng Zhang}
\affiliation{%
  \institution{Northeastern Univ., China} 
  % \streetaddress{}
  % \city{Shenyang}
  % \state{China}
  % \postcode{110000}
}
\email{zhangyf@mail.neu.edu.cn}

\author{Feng Yao}
\affiliation{%
  \institution{Northeastern Univ., China} 
  % \streetaddress{}
  % \city{Shenyang}
  % \state{China}
  % \postcode{110000}
}
\email{yaofeng@stumail.neu.edu.cn}

\author{Wenyuan Yu}
\affiliation{%
  \institution{Tongyi Lab, Alibaba Group}
}
\email{wenyuan.ywy@alibaba-inc.com}

\author{Ge Yu}
\affiliation{%
  \institution{Northeastern Univ., China}  
}
\email{yuge@mail.neu.edu.cn}

\author{Jingren Zhou}
\affiliation{%
  \institution{Tongyi Lab, Alibaba Group}
}
\email{jingren.zhou@alibaba-inc.com}

% \orcid{0000-0002-1825-0097}
% \affiliation{%
%   \country{Iceland}

% \orcid{0000-0001-5109-3700}
% \affiliation{%
% \email{vb@rocquencourt.com}

% \author{J\"org von \"Arbach}
% \affiliation{%
%   \institution{University of T\"ubingen}
%   \city{T\"ubingen}
%   \country{Germany}
% }
% \email{jaerbach@uni-tuebingen.edu}
% \email{myprivate@email.com}
% \email{second@affiliation.mail}

% \author{Wang Xiu Ying}
% \author{Zhe Zuo}
% \affiliation{%
%   \institution{East China Normal University}
%   \city{Shanghai}
%   \country{China}
% }
% \email{firstname.lastname@ecnu.edu.cn}

% \author{Donald Fauntleroy Duck}
% \affiliation{%
%   \institution{Scientific Writing Academy}
%   \city{Duckburg}
%   \country{Calisota}
% }
% \affiliation{%
%   \institution{Donald's Second Affiliation}
%   \city{City}
%   \country{country}
% }
% \email{donald@swa.edu}

\begin{abstract}
% There have been growing demands in dynamic graph processing, in which a continuous stream of graph updates is mixed with graph computation.
An efficient data structure is fundamental to meeting the growing demands in dynamic graph processing.
However, the dual requirements for graph computation efficiency (with contiguous structures) and graph update efficiency (with linked list-like structures) present a conflict in the design principles of graph structures.
%, creating a complex challenge in the design of graph storage suitable for dynamic graph environments.
%Merely adjusting the continuity of the data structure is insufficient to simultaneously improve the performance of both.
After experimental studies of state-of-the-art dynamic graph structures, we observe that the overhead of cache misses accounts for a major portion of the graph computation time. 
%Based on this observation, we aim to improve the efficiency of both computation and updating by reducing the overhead of cache misses.
%our approach
This paper presents \oursys, a system with graph storage and coroutine-based prefetch co-design. %, which is designed to improve computational efficiency and update support in dynamic graph processing.
By employing software prefetching via stackless coroutines and designing a prefetch-friendly data structure \ourstorage, \oursys significantly alleviates the performance degradation caused by cache misses. 
%\todo{Our results show that \oursys outperforms state-of-the-art systems by 1.3$\times$  - 180$\times$ in data insertion and on average, 11.6$\times$ in graph analysis algorithms.}
%\ys{
Our results show that \oursys outperforms state-of-the-art graph storage systems by 1.3$\times$  - 180$\times$ in graph updates and 1.4$\times$ - 41.1$\times$ in graph computation.
%}

% Our results indicate that \oursys outperforms state-of-the-arts graph storage by a factor of 1.3~180\times in data insertion, and it is, on average, 11.6 times better in graph analysis algorithms.
% 我们的结果显示\oursys在数据插入上优于最先进的图存储1.3~180\time,并且在图分析算法上平均优于11.6\time.

% Finally, the experiments show that RAGraph can achieve 1.69× - 40.53× speedup and 20.9% - 97% WAN cost reduction compared with state- of-the-art systems

%our exp
%We also conduct extensive experiments on different datasets and show that \todo{\oursys{} could outperform state-of-the-arts by 10.48× on average and at the same time guarantee a pioneer insertion performance (1st place in 5 cases and 2nd place in 2 cases).}
\end{abstract}

\maketitle

\pagestyle{\vldbpagestyle}

% \eat{
\begingroup\small\noindent\raggedright\textbf{PVLDB Reference Format:}\\
\eat{\vldbauthors}
Hongfu Li, Qian Tao, Song Yu, Shufeng Gong, Yanfeng Zhang, Feng Yao, Wenyuan Yu, Ge Yu, and Jingren Zhou. \vldbtitle. PVLDB, \vldbvolume(\vldbissue): \vldbpages, \vldbyear.\\
\href{https://doi.org/\vldbdoi}{doi:\vldbdoi}
\endgroup
\begingroup
\renewcommand\thefootnote{}\footnote{\noindent
This work is licensed under the Creative Commons BY-NC-ND 4.0 International License. Visit \url{https://creativecommons.org/licenses/by-nc-nd/4.0/} to view a copy of this license. For any use beyond those covered by this license, obtain permission by emailing \href{mailto:info@vldb.org}{info@vldb.org}. Copyright is held by the owner/author(s). Publication rights licensed to the VLDB Endowment. \\
\raggedright Proceedings of the VLDB Endowment, Vol. \vldbvolume, No. \vldbissue\ %
ISSN 2150-8097. \\
\href{https://doi.org/\vldbdoi}{doi:\vldbdoi} \\
}\addtocounter{footnote}{-1}\endgroup

\ifdefempty{\vldbavailabilityurl}{}{
\vspace{.3cm}
\begingroup\small\noindent\raggedright\textbf{PVLDB Artifact Availability:}\\
The source code, data, and/or other artifacts have been made available at \url{https://github.com/GorgeouszzZ/GastCoCo}.
\endgroup
}
\renewcommand{\thefootnote}{\fnsymbol{footnote}}
\footnotetext[1]{Hongfu Li and Qian Tao contributed equally.}
% \footnotetext[2]{Yanfeng Zhang is the corresponding author.}
\renewcommand{\thefootnote}{\arabic{footnote}}

\section{Introduction}
\label{sec:intro}

% \begin{figure}
%   \centering
%   \includegraphics[width=\linewidth]{fig-sec1/RW.pdf}
%   \caption{An e-commerce purchase graph and the memory storage of real data.}
%   \label{fig:real-world-eg}
% \end{figure}

\begin{figure}[ht]
  \centering
    \includegraphics[width=0.94\linewidth,trim=0 202 0 0,clip]{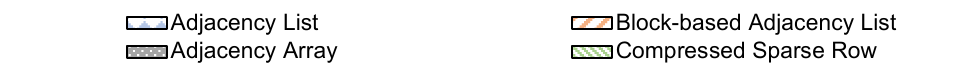}
    \vspace{-0.14in}
    \\
    \subfloat[Algorithm execution time and graph update time.]{\label{fig:alg-time}
    \includegraphics[width=0.94\linewidth,trim=0 0 0 45,clip]{fig-sec1/alg-2.pdf}}
    \\
    \vspace{-0.1in}
    % \subfloat[CPU cache stall proportion. ]{\label{fig:cache-miss-percent}
    % \includegraphics[width=0.94\linewidth]{fig-sec1/alg-cm-2.pdf}} 
    \subfloat[CPU cache stall count. ]{\label{fig:cache-miss-percent}
    \includegraphics[width=0.94\linewidth]{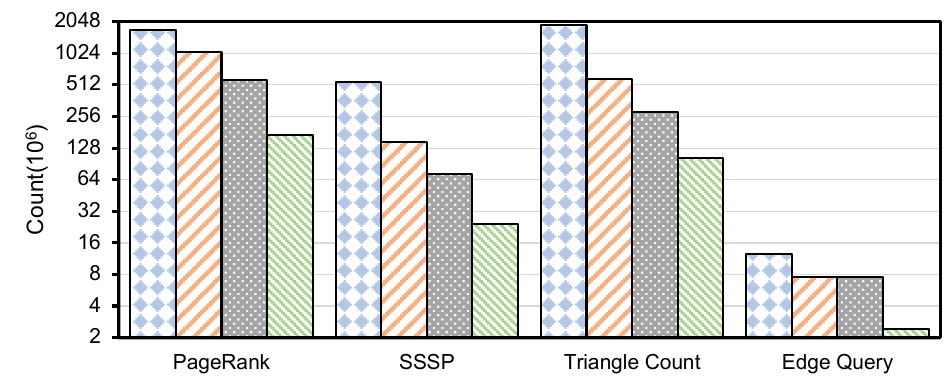}} 
    \vspace{-0.1in}
  % \caption{The execution time for graph algorithms (and graph updates) and the CPU cache stall proportion on different data structures. T.O. means that the graph updates cannot be finished in a reasonable time (much longer than 48 hours).}
  \caption{The execution time for graph algorithms (and graph updates) and the CPU cache stall count on different data structures (T.O.: graph updates cannot finish in 24 hours).}
  %\zyf{color label is not obvious. from fig b, csr is always best. should show the bad performance of csr for insertion, otherwise why not use csr? }}
  \label{fig:alg-and-cm}
  \vspace{-0.2in}
\end{figure}

With the increase in data scale and diversification of data types, dynamic graph processing has become a critically important issue in various domains such as e-commerce, financial technology, and social networks~\cite{DBLP:journals/cbm/AzadifarRBMO22, DBLP:conf/cikm/TuQWZLZWZZ23, DBLP:journals/information/ZaminiRR22}.
Specifically, dynamic graph processing involves executing graph computation algorithms on graphs that experience frequent structural changes, such as millions of edge updates per hour~\cite{DBLP:journals/pvldb/SahuMSLO17, he2023graphscope}.
For instance, there are more than $400$ million behaviors per day~\cite{AlibabaDAU} between users and items on Taobao~\cite{Taobao}. The platform should employ a combination of graph analytics and interactive graph traversals for fraud detection or interactive pattern mining~\cite{BolinDingKaiZengWenyuanYu}. 
Such applications on dynamic graphs require not only efficient graph computations but also supporting the large volume of updates per second~\cite{DBLP:journals/pvldb/GuptaSGGZLL14, DBLP:journals/pvldb/QiuCQPZLZ18, DBLP:journals/pvldb/SharmaJBLL16}.

Regrettably, the need for computation efficiency and frequent update support presents a contradictory challenge, leading to a dilemma in designing storage structures for dynamic graphs.
Take the widely used graph storage structures, Adjacency List (AL) and Compressed Sparse Row (CSR) ~\cite{DBLP:books/daglib/0009092}, as an example.
\eat{crc: AL and CSR are two extremes in graph storage: CSR is completely contiguous in memory, while AL is highly discontiguous in memory.}
AL is a graph data structure where the neighbors of each vertex are stored in a linked list.
Since the nodes in the linked list are non-contiguous memory fragments connected by pointers, it is easy to insert/delete nodes without data movements on the AL structure, which is suitable for graph updates.
On the other hand, 
% For example, Compressed Sparse Row (CSR) ~\cite{DBLP:books/daglib/0009092} is a widely used graph storage structure in static graphs because it is not only memory efficient but also graph computation efficient. 
CSR has been widely used for graph storage in static graphs due to its memory and computational efficiency.
Figure\autoref{fig:alg-time} reports the execution time for graph computations (PageRank, SSSP, Triangle Count, and Edge Query) and updates (Insertion) under a single thread on the LiveJournal dataset~\cite{Livejournal_dataset}.
We could observe that CSR is more efficient than AL on graph algorithms, but spends more time for updating graphs.\eat{crc: (Graph update experiment on CSR cannot finish in 48 hours).}

% In summary, Graph storage stored in highly contiguous memory (to an extreme, CSR) can lead to a more efficient graph computation \gsf{but with poor graph update performance}, while fully fragmented memory storage may be beneficial for the updates of graphs (to an extreme, AL) \gsf{but has poor graph computation performance}.
% \ys{Change the order of describing AL and CSR?} \lhf{finish.} \zyf{This paragraph is not a novel observation, the fact is not insight, reviewers should know it. compact to a sentence and add it to the end of last paragraph.}

%sketch:CSR和AL在支持图计算和图更新上两个极端，因为整个图在CSR中连续，而AL分散。CSR的连续有利于图计算过程中cache miss，但是不利于更新。而AL的分散有利于更新，但存在大量cache miss。如图X所示。为了能同时做得更好点，可以采用折衷方法，例如AA，BAL，【简介AA，BAL】。最近出现一些更巧妙的设计grahponeXXX，他们根据图的一些特性进行了更细致的设计，比如顶点度数的分布等等。但是上述方法均是顾此失彼，要么通过连续性，减少cache miss，但是更新不好，要么更分散，更新性好，但是cache miss增多。
%另起一段：解决顾此失彼的方法有两种，一种是仍然连续，同时支持快速更新，livegraph，但是存在XX问题。另一种是在不连续的基础上减少cache miss。
To effectively support both graph computation and graph updates, some efforts propose designing graph structures that maintain a moderate degree of contiguity~\cite{zhu2019livegraph, fuchs2022sortledton, pandey2021terrace, ediger2012stinger, DBLP:journals/tos/KumarH20}.
% A straightforward approach would be to use a halfway house between CSR and AL, which means adjusting the degree of
% contiguity of the data structures.
For example, Adjacency Array (AA)~\cite{zhu2019livegraph} stores all neighbors of each vertex in a separate array and Block-based Adjacency List (BAL)~\cite{ediger2012stinger} stores 
the neighbors of each vertex in a block-based linked list (\ie a node contains more than one neighbors%a linked list composed of blocks with each block storing multiple neighbor information
).
\eat{crc:are such compromised storage structures, where AA stores all neighbors of each vertex in a separate array and BAL stores 
the neighbors of each vertex in a block-based linked list (\ie a node contains more than one neighbors%a linked list composed of blocks with each block storing multiple neighbor information
).}
The degrees of contiguity in these two structures fall between CSR and AL.
More recently, more complex and detailed implementations, 
\eg the state-of-the-art dynamic graph storage systems like GraphOne~\cite{DBLP:journals/tos/KumarH20}, Stinger~\cite{ediger2012stinger}, Terrace~\cite{pandey2021terrace}, and Sortledton~\cite{fuchs2022sortledton}, maintain the graphs with a linked data structure fragmented in memory with various degrees of contiguity \eat{crc:(more akin to CSR or AL) }as a halfway house. 
However, such a halfway-house solution can only balance the performance of graph computation and graph updates, which sacrifices the efficiency of graph computation to enhance graph update throughput, or vice versa.
To avoid differences in optimization and implementation across systems, we use uniformly implemented simple data structures for the cache stall count experiments in Figure~\ref{fig:cache-miss-percent}.
As representatives of ``halfway-house solutions'', BAL and AL exhibit a large number of cache misses in graph computation tasks.
% As representatives of "a halfway-house solution," BAL and AL exhibit a significant proportion ($71.38\%$ to $96.21\%$) of cache miss overhead in graph computation tasks.% /------------response---------------
% Alternatively, we observe that the decline in graph algorithm performance from CSR to AL is derived from the increase in the number of cache misses caused by the pointer chasing~\cite{DBLP:conf/asplos/LukM96, he2020corobase}.
% % For the second method, 
% We find that massive cache misses occur, particularly with linked data structures, when fetching data from memory across all algorithms.
% \autoref{fig:alg-and-cm} shows trends of execution time and proportion of CPU cache stall time when executing computation on the structures CSR, AA, BAL, and AL.
% The overhead caused by cache misses constitutes a major part ($50.22\%-97.72\%$) of the execution time, and as the degree of discontinuity of memory increases, the CPU cache stall time increases.
% % \lhf{This is because the less contiguous the data is in memory, the worse its spatial locality, leading to poorer cache locality.}
% This is because the more discontiguous data in memory, the worse the spatial locality, which results in less effective cache utilization.
% % We observe that the frequent cache misses are actually the main component of the running time and the more part of discontinuous memory \zyf{component and the more part are misused, don't understand}, the larger the proportion of the cache miss time \zyf{wrong sentence, non-sence, no one use it like this}.
% /------------response---------------
This motivates us to seek a novel perspective to improve both the efficiency and dynamic graph support of dynamic graph processing:
% beyond the contradiction for dynamic graph processing.
\emph{Can we mitigate the overhead of cache misses from the linked data structures to improve the efficiency for both computation and updating?}

To answer the question, we consider a co-design of the prefetching techniques and graph structures.
Prefetching techniques allow us to load data, which will be accessed later, from memory into the CPU cache, thereby reducing the data access latency.
These techniques consist of both hardware prefetching, where the CPU automatically loads data and instructions, and software prefetching, where programmers explicitly insert instructions into the code to prefetch data.
Although hardware prefetching techniques are commonly equipped on standard hardware, software prefetching techniques have recently demonstrated their effectiveness in many applications ~\cite{callahan1991software, he2020corobase, corograph}.
% We can leverage both hardware prefetching and software prefetching techniques to reduce the cache miss overhead  \gsf{what's the overhead caused by cache miss?}, thereby enhancing the performance of both graph computation and graph updates. % 这句话没什么营养，基本上跟上一句话的意思一样
However, when it comes to the practical dynamic graph processing domain, there are still challenges. 
% But in the practical dynamic graph processing domain, there are three \textbf{challenges}:
% \ding{172}How to enhance the performance of hardware prefetching?
% \ding{173}What are the methods and timing for using software prefetching?
% \ding{174}How can hardware prefetching and software prefetching be better coordinated?

\noindent
\textbf{Our approach.}
This paper presents \oursys, an in-memory system with \underline{G}r\underline{A}ph \underline{ST}orage and \underline{CO}routine-based prefetch \underline{CO}-design, which is designed for dynamic graph processing applications to alleviate the performance overhead caused by cache misses.
% \oursys shows performance improvements in virtually all dynamic graph processing applications.
%Prefetching可以提前将程序需要的数据加载到CPU cache中，避免cache miss latency。包括硬件预取和软件预取。然后再说我们用了这个玩意。
% We use prefetching techniques to load data that is likely to be accessed soon from memory into the CPU cache to reduce access latency.
% \oursys leverages both hardware prefetching and software prefetching techniques to reduce the overhead caused by cache misses \gsf{what's the overhead caused by cache miss?}, thereby enhancing the performance of both graph computation and graph updates.
\ding{172}In terms of hardware prefetching, the prefetching effectiveness varies across different data structures.
\oursys proposes a prefetch-aware data structure \ourstorage for dynamic graph processing.
\ding{173}Software prefetching is used to compensate for cases where hardware prefetching fails, so the benefits of applying software prefetching are our focus.
\oursys develops a set of hybrid prefetching strategies to avoid fetching reduplicated data by both software and hardware prefetching.
To minimize the overhead of applying software prefetching and increase the prefetch success rate, \oursys leverages C++20~\cite{C++20} stackless coroutines to prefetch the graph data that the program requires to access, benefiting both graph computation and graph updates by mitigating the cache miss overhead.
\ding{174}Finally, \oursys designs a set of adaptive coroutine scheduling and task allocation strategies tailored to different graph tasks, allowing graph processing using software prefetching to perform better on \oursys.
% \todo{Compared with previous state-of-the-art systems, the proposed system achieves the $10.48\times$ speedup on average in terms of algorithms like PageRank and SSSP, and $1.6$ rank on average in terms of the throughput for graph insertions.}

\noindent
\textbf{Contributions.}
To sum up, the contributions of this paper include:
\begin{itemize}[leftmargin=*]
    % \vspace{-0.4in}
  \item The first to employ the instruction stream interleaving execution mode composed of coroutines and software prefetching to reduce cache misses in dynamic graph processing, thereby enhancing the performance of both graph computation and graph updates.
  \item A dynamic graph data structure \ourstorage, specifically designed for efficient dynamic graph processing, which not only improves the effectiveness of hardware prefetching but also facilitates the implementation of software prefetching via coroutines.
  % \item A hybrid prefetching strategy that effectively utilizes both hardware and software prefetching to enhance performance, enabling it to adapt to a wide range of environments.
  \item An efficient graph storage system \oursys that is equipped with a set of optimizations, \eg %A comprehensive strategy \zyf{use specific name, like xxx system GastCoCo that leverages ..... and supports a set of optimizations}, including 
  task allocation, coroutine scheduling, and hybrid prefetching, to minimize the overhead of software prefetching via coroutine as much as possible on different graph tasks and runtime environments.
Our results show that \oursys outperforms state-of-the-art graph storage systems by 1.3$\times$  - 180$\times$ in graph updates and 1.4$\times$ - 41.1$\times$ in graph computation.
%  \item \todo{show experiment.}
\end{itemize}

\section{Preliminaries}
In this section, we first summarize the data access patterns in graph computations and then discuss how hardware and software prefetching work in graph processing.

\subsection{Graph Operations and Data Access Patterns}
\label{subsec:acc_pattern}

There are numerous types of dynamic graph processing tasks and their performances are usually regarded as memory access bounded \cite{DBLP:conf/wosp/EisenmanCMCFK16, DBLP:conf/iiswc/BeamerAP15}.
Most of these tasks can be deconstructed into a bunch of data access operations on vertices and edges. %(besides this, there are operations and computations on the data, we don't discuss this part in this section).
%By thoroughly analyzing these data access operations, we can extrapolate the analysis results to all dynamic graph processing tasks.
%zyf: 所有图计算的IO pattern可以归纳为5个基本操作，只有读？是不是I/O pattern。这句话好好查查怎么写，我也不会。
%Additionally, data access operations enable us to obtain the predictability of the data about to be accessed, allowing us to implement more accurate and efficient prefetching strategies.
%So we summarize the following data access operations on vertices and edges from dynamic graph processing tasks.
The operations to access vertices are listed as follows.
\begin{itemize}[leftmargin=*]
    \item \PrAllVs traverses all vertices in the entire graph. 
    \item \PrAllVsCond makes certain conditional filtering during the traversal, and the vertices that meet the conditions $cond$ can undergo subsequent operations.
    \item \PrOneV reads a specific vertex $v$.
\end{itemize}
Processing edges can be considered as first locating the edges via the source vertex $v_{src}$ and then processing the neighbors of $v_{src}$.
Since locating the source vertex can be achieved by \PrOneV, our focus is exclusively on the process of neighbors.
\begin{itemize}[leftmargin=*]
    \item \PrAllNeis traverses all neighbors of a vertex $v_{src}$. 
    %$v_{src}$ is a parameter signifying the source vertex in such an operation, which means reading all adjacent edges of $v_{src}$.
    \item \PrOneNei reads a particular edge from a source vertex $v_{src}$ to a destination vertex $v_{dst}$.
\end{itemize}
For example, in each iteration, Single Source Shortest Path (SSSP) algorithm ~\cite{DBLP:journals/tpds/ChakaravarthyCM17} requires processing all neighbors of vertices whose status was updated in the previous iteration. 
This can be represented as a combination of \PrAllVsCond and \PrAllNeis. 
The edge query~\cite{DBLP:journals/pvldb/SahuMSLO17} between two vertices involves \PrOneV and \PrOneNei operations.

In summary, these operations can be categorized as sequential and random memory access.

\etitle{Sequential Data Access.}
Some tasks require sequential data access, \ie to access the storage structure in a fixed order (\eg the lexicographical order).
If the structure is physically contiguous in memory, sequential data access can be regarded as sequential memory access.
For example, \PrAllVs and \PrAllNeis can be regarded as sequential data access or memory access on arrays.

\etitle{Random Data Access.}
Random data access refers to accessing the storage structure (or part) in a random order. 
Since the access is unpredicted, performing random data access on either contiguous or non-contiguous memory can be considered as random memory access.
Correspondingly, \PrAllVsCond can be regarded as random data access because $cond$ is unpredictable.
\PrOneV and \PrOneNei also access data in a random order.

% Different graph data access operations involve different access patterns.
% For \PrAllVs and \PrAllNeis, they are sequential access. 
% % When the data to be accessed are arranged contiguously in memory, these operations can achieve the best performance due to cache locality (\eg CSR).
% For \PrAllVsCond, due to the unpredictable feature of the condition $cond$, this operation is almost random access.
% % In addition, it is worth noting that to obtain efficient support for graph pattern matching algorithms, we require the data structures to maintain the order of vertices and edges. 
% For \PrOneV and \PrOneNei, they access randomly according to a specific vertex $v$ or a specific edge $(v_{src}, v_{dst})$.
%we choose more efficient search strategies, such as binary search instead of traversal search when searching in an array. Therefore, we consider they randomly access memory.

\begin{figure}[t]
  \centering
  \includegraphics[width=0.95\linewidth]{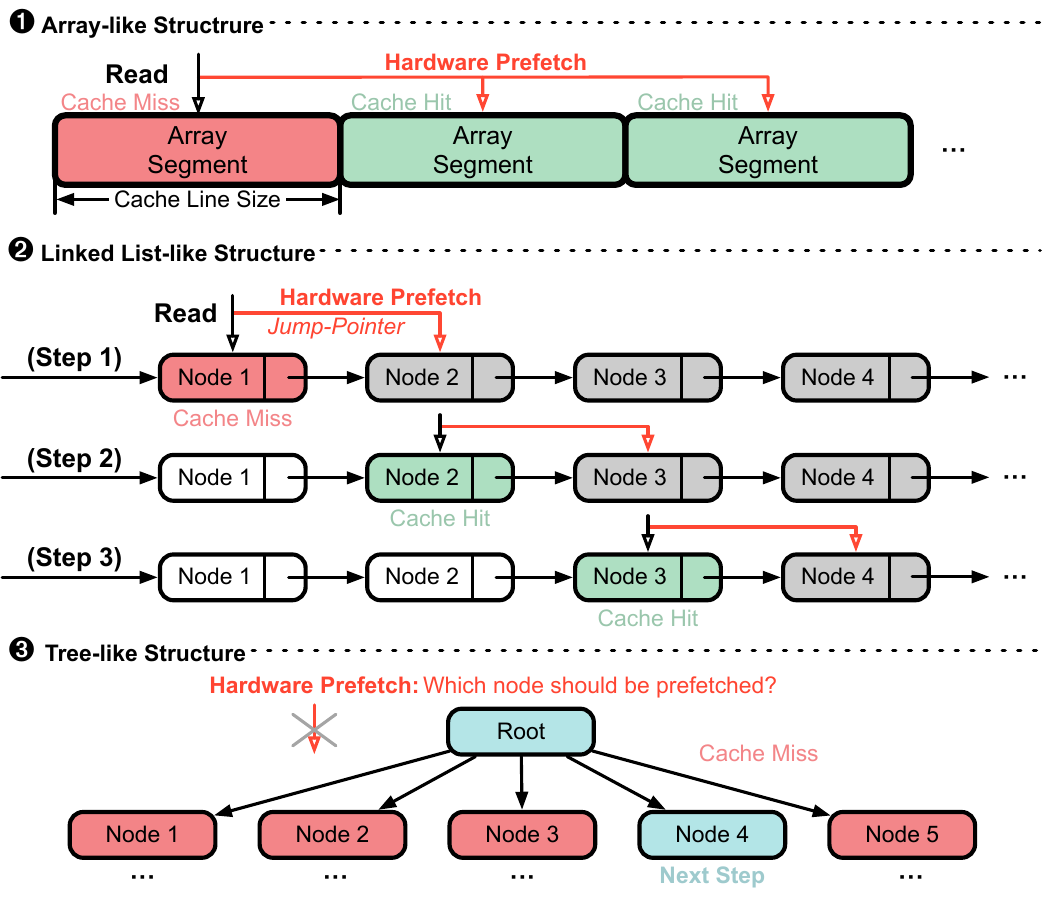}
  \vspace{-0.15in}
  \caption{Hardware prefetching.% (Jump-pointer mechanism prefetches 1 Node ahead \zyf{what's this? remove})
  }
  \label{fig:JPM}
  \vspace{-0.1in}
\end{figure}

\subsection{Hardware Prefetching in Graph Processing}
% \subsection{\tao{Cache Misses Occupy Time Cost}} %Hardware Prefetching in Graph}
\label{subsec:cache_miss_and_hard_pre}
Hardware prefetching is a predefined mechanism in the processor that, based on the data stream requested by the running program, identifies and prefetches the subsequent elements that the program might need, loading them into the processor's cache beforehand~\cite{DBLP:conf/sc/BaerC91}.
However, its effectiveness is affected by different data structures.
Dynamic graph storage generally incorporates contiguous and non-contiguous memory structures to support efficient graph computation and updates. 
Specifically, Array-like structures~\cite{DBLP:books/daglib/0009092, DBLP:conf/hpec/WheatmanX18, de2021teseo} use contiguous memory, while Linked List-like~\cite{fuchs2022sortledton, DBLP:journals/tos/KumarH20, ediger2012stinger} and Tree-like structures~\cite{dhulipala2019low, pandey2021terrace, fuchs2022sortledton} use non-contiguous memory.

% In general, there are three storage structures to store the dynamic graphs, CSR~\cite{}, BAL~\cite{}, and tree~\cite{}. 
%Next, we will introduce the role of hardware prefetching in enhancing access to these structures.

% \todo{may merge.}
\etitle{Hardware prefetching in Array-like structures.}
The data in an Array-like structure is stored contiguously in the main memory.
As shown in \autoref{fig:JPM} \blackcircleone, when \PrAllNeis is performed on an array, the hardware prefetchers prefetch the next several segments adjacent to the requested segment into the cache from the main memory.
Here, the size of a segment matches that of a cache line \eat{(segment size matches cache line size)}~\cite{DBLP:journals/computer/Smith78, DBLP:conf/sc/BaerC91, DBLP:series/synthesis/2014Falsafi}.
% \gsf{ when loading XX from memory to cache, the XX also be prefetched from memory to cache by the way.}
%As common sense, one mechanism of hardware prefetching is to prefetch data in adjacent cache lines~\cite{DBLP:journals/computer/Smith78, DBLP:conf/sc/BaerC91, DBLP:series/synthesis/2014Falsafi}, \gsf{Do you copy the original sentence? add an illustrated example} 
Therefore, when performing sequential memory access on array-like structures (\eg \PrAllNeis in CSR), most of the required data will be hit in the CPU cache. %most data can hit the cache. 
% However, the Array-like structures require massive data movement when inserting new edges, \eg CSR. 
% The data movement caused by inserting edges in an Array-like structure can be alleviated by reserving some empty cells in the Array, \eg PMA ~\cite{DBLP:journals/tods/BenderH07}, but the empty cells result in a waste of memory.
However, if the operation randomly accesses data, most of the data will be missed.

\etitle{Hardware prefetching in Linked List-like structures.}
\eat{remove: As discussed in \autoref{sec:intro}, most previous works employ structures that use non-contiguous memory for fast graph updates. 
% In this subsection, we move a step forward to demonstrate how pointer chasing incurs cache misses during sequential access in non-contiguous memory~\cite{he2020corobase, jonathan2018exploiting, DBLP:journals/pvldb/PsaropoulosLMA17, DBLP:conf/nsdi/FanAK13, DBLP:conf/asplos/LukM96}. \gsf{gap, see comment}%gsf:最后这一句开始说要解释ponter chase如何造成cache miss，后边怎么还又说不连续内存的顺序访问？弄不明白你到底要说啥。而下边这段说的是hardware prefetching。感觉没任何联系啊。
Specifically, many of these dynamic graph storage~\cite{fuchs2022sortledton, DBLP:journals/tos/KumarH20, ediger2012stinger}employ List-like structures.}
The data stored in the Linked List-like structure is discontiguous, and pointers in memory connect the data.
Under the sequential data access pattern, the data to be accessed next can be known in advance via the fixed access paths.
Modern hardware prefetchers prefetch data in the sequential data access pattern by \emph{jump-pointer mechanism}~\cite{DBLP:conf/asplos/LukM96,  DBLP:conf/micro/CollinsSCT02, Roth_Moshovos_Sohi_1998, DBLP:conf/isca/RothS99, DBLP:series/synthesis/2014Falsafi}.
In \autoref{fig:JPM} \blackcircletwo, the jump-pointer mechanism prefetches Node $2$ when the processor reads Node $1$, preventing a cache miss when the processor reads Node $2$.
However, such a jump-pointer mechanism does not work for the head of a linked list.
% for Jump Pointer mechanism to be effective, prefetching must occur in advance.
For instance, upon initially entering the linked list in \autoref{fig:JPM} \blackcircletwo, Node $1$ can not be prefetched, leading to a cache miss.
We offer the following explanation regarding the occurrence of cache misses:
% i) The longer duration for \PrAllNeis on the first block is because the first block is not located in the caches with a high possibility when initially accessing a new block-based linked list.
% ii) Even if the memory address of the next block is far from the current block, modern hardware prefetchers can still shorten the duration of \PrAllNeis on subsequent blocks because of the jump-pointer mechanism designed for pointer-chasing access patterns.
% The jump-pointer mechanism can load the afterward blocks in advance while walking on a linked list as shown in \autoref{fig:JPM}, thus is particularly effective for traversing linked lists~\cite{DBLP:series/synthesis/2014Falsafi}.
i) The first node is not located in the caches with a high possibility while accessing a linked list.
Therefore, it results in a cache miss (in fact, the front few nodes might trigger cache misses, depending on how many nodes the jump-pointer mechanism prefetches ahead).
ii) Even if the memory address of the next node is far from the current block, the jump-pointer mechanism can load the afterward nodes in advance while traversing on a linked list as shown in \autoref{fig:JPM}. 
Therefore, fewer cache misses occur in the subsequent process.

\etitle{Hardware prefetching in Tree-like structures.}
Several previous works, like Aspen~\cite{dhulipala2019low} and Terrace~\cite{pandey2021terrace}, utilize tree-like structures to store the\eat{graphs for} high-degree vertices.
In these structures, 
% \PrAllNeis will shift from sequential access to random access.
there is little inherent spatial locality between accessed nodes since they are dynamically allocated from the heap and can have arbitrary addresses~\cite{DBLP:conf/asplos/LukM96}.
The hardware prefetching no longer works for such structures~\cite{he2020corobase, DBLP:series/synthesis/2014Falsafi} as blocks may contain more than one pointer, and hardware prefetchers cannot predict the path that will be selected.
Therefore, both \PrAllNeis and \PrOneNei (sequential and random data access in a tree) result in a large number of cache misses.

In summary, hardware prefetching mechanisms provide more direction for our data structure design. 
However, the unexpected performance of hardware prefetching inspires us to employ other modern techniques to \eat{reduce the time of cache misses or even }eliminate some of the cache misses from data fetching, such that a fragmented data structure for dynamic graph computing could also achieve better performance.

\subsection{Software Prefetching via Coroutines}
\label{subsec:coro_SP}
% \gsf{see comment}%别上来就说以前的工作怎么怎么滴。先说coroutine是什么，可以用来数据获取减少cache miss。该方法已经用在数据库和图数据处理。
Although hardware prefetching relieves the access time for contiguous data, it performs badly on graphs.
In response, recent efforts utilize coroutines~\cite{moura2009revisiting, jonathan2018exploiting} to mitigate cache misses.
% \lhf{We hope software prefetching can compensate for cases where hardware prefetching fails.}
% \todo{<-sPlease Dr.Tao modify it.}
For example, prior works~\cite{jonathan2018exploiting, chen2007improving, DBLP:journals/pvldb/PsaropoulosLMA17, he2020corobase} propose optimizing the overhead of cache misses caused by random memory access, by introducing software prefetching via coroutines.
Software prefetching techniques \cite{jonathan2018exploiting, chen2007improving, DBLP:journals/pvldb/PsaropoulosLMA17, kocberber2015asynchronous} leverage workload semantics to issue prefetch instructions~\cite{Intel2016} to explicitly bring data into CPU caches~\cite{he2020corobase}.

\eat{Unlike hardware prefetching, software prefetching is controlled by the instructions written by programmers explicitly.}%accomplished by programmers manually setting instructions and determining the content to be prefetched.

This solution enables the overlapping of data loading and computation.
Data can be explicitly loaded using prefetch instructions as needed.
However, given the delay involved in moving data from memory to cache, the CPU switches to another task to compute to overlap with data loading time.
We need a tool to finish the switching between tasks.
While the multithreading approach is straightforward, it has much heavy overhead on thread switching relative to the cost of LLC cache miss.
To this end, we need new techniques to switch between tasks.
% Our original intention was to reduce the overhead of cache misses through this solution.
% Opting for multithreading would result in a loss rather than a gain.

% Coroutines are functions that can be suspended and resumed by the programmers.
% \gsf{see comment}%后边对coroutine的描述是不是可以挪到这？最后一句看不出software prefetching是怎么工作的。是不是这么个逻辑：当需要读取数据时，可以将当前协程suspend，切换到其他协程，其他协程完毕后，再resume该协程，此时数据已经获取？

% This solution enables the overlapping of data loading time with computation time:
% Data access tasks with pointer chasing are written into coroutines.
% During execution, each coroutine, upon encountering the dereferencing of a pointer \gsf{it's hard to understand}, actively issues a software prefetch request followed by suspending itself and switches to another coroutine~\cite{he2020corobase} \gsf{you mean more than one coroutines in a thread? Please point it first}. 

\etitle{C++20 Stackless Coroutine}
% taoqian
% 这里是分析，首先简要介绍coroutine的历史，然后需要给读者介绍一个大致的概念，为什么coroutine可以用来“聪明地”解决图计算中的预取问题
% lhf: 第二个问题为什么“聪明地”解决？没想好怎么回答？
are special functions that can be suspended voluntarily and resumed later at low cost during the execution~\cite{moura2009revisiting, conway1963design, he2020corobase}. 
% /-----------------for response--------------
The efficient suspension and resumption of stackless coroutines make it possible to implement the interleaving execution mode.
Specifically, when issuing a software prefetching instruction, we can suspend the coroutine, switch to another coroutine for computing, and later resume the previous coroutine to continue the corresponding task.
% /-----------------for response--------------
% 关于stackless和stackful的讨论感觉没必要单独讨论，跟在c++ stackless coroutines后面讨论下stackless的优势就可以了
% Coroutines can be classified into two types: stackful coroutines~\cite{moura2009revisiting} and stackless coroutines. 
% The term "stackful" refers to the presence of a separate runtime stack for each coroutine. 
% As suspension and resumption are involved, it is necessary to save the context. 
% Stackful coroutines, such as the widely used "libco library" in WeChat, employ assembly language instructions to store the context (i.e., the values of each register variable) in the call stack.
% On the other hand, 
% /-----------------------for response-----------------------------
Stackless coroutines are essentially state machines, \ie coroutine switching simply involves changing the instruction pointer register. 
All coroutines within a physical thread share the system stack, and it is unnecessary to save the states of registers explicitly.
% % To conclude, stackless coroutines have lower overhead in context switching compared to stackful coroutines.
The stackless coroutines in C++20~\cite{C++20} standard exhibit low overhead in terms of construction and context switching\eat{ (cheaper than a last-level cache miss)}~\cite{he2020corobase}, and \oursys is constructed based on C++20 stackless coroutines.
% /-----------------------for response-----------------------------

% “从challenge得到opportunity”的写法看起来有一点本末倒置，意味着“因为有challenge所以要做”。更好的表现opportunity的方式应该是有一些观察等驱动我们做图上的software coroutine。这里建议两段合成一段，然后通过figure 1说明我们可以通过减少cache miss的时间来突破robbing peter to paul的限制
% This solution, which reduces cache miss overhead through software prefetching of data, still presents numerous challenges:
% i) Since this solution primarily models coroutines as transactions in databases, how should it be applied in dynamic graph processing domain? \gsf{corograph?}
% ii) There are certain costs associated with this solution \gsf{then?}.
% In cases of software prefetching failure or redundant prefetching of content already in the cache, how should these issues be resolved?
%1、本身有开销（什么开销？），如果已经在cache则会脱裤子放屁吧。2、硬件预取会预取，如何配合硬件预取？

Hardware prefetching is an automatic prefetching technique executed by the processors but shows inefficiency in many cases (as discussed in \autoref{subsec:cache_miss_and_hard_pre}).
We apply software prefetching to complement hardware prefetching, achieving a more effective performance.
Switching between coroutines can cause considerable cost, and our work focuses on how to implement effective software prefetching for dynamic graphs.
\section{Overview of \oursys}

\begin{figure}[t]
  \centering
  \includegraphics[width=0.9\linewidth]{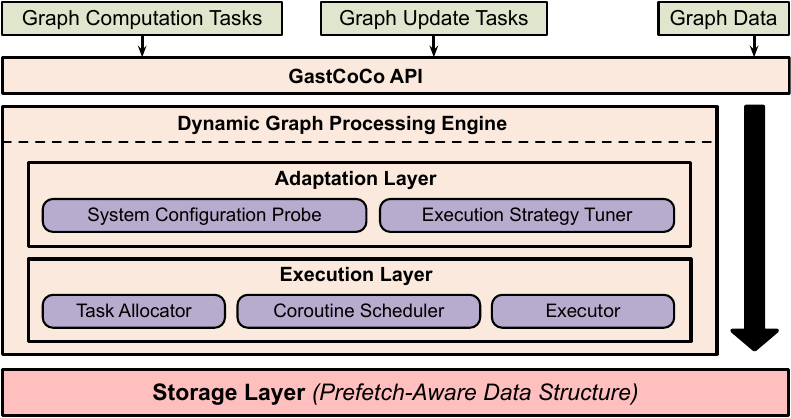}
  \caption{Overview of \oursys.}
  \label{fig:gastcoco}
  \vspace{-0.15in}
\end{figure}

\eat{ % old version
\begin{table}[h]
  \vspace{-0.1in}
    \caption{\lhf{\oursys APIs.}}
    \label{tab:api}
    \begin{tabular}[t]{rlll}
        \toprule
        Name & Parameters & & Response \\
        \midrule
        \textbf{LoadGraph} & (file\_path) & $\rightarrow$ & load\_state \\
        \textbf{UpdateVertex} & (obj\_v, update\_mode) & $\rightarrow$ & update\_state \\
        \textbf{UpdateEdge} & (obj\_e, update\_mode) & $\rightarrow$ & update\_state \\
        \textbf{BatchUpdate} & (update\_content) & $\rightarrow$ & update\_state \\
        \midrule
        \textbf{ProcessVertex} & (f, active) & $\rightarrow$ & result \\
        \textbf{ProcessEdge} & (dense\_f, sparse\_f, active) & $\rightarrow$ & result \\
    \bottomrule
    \end{tabular}
\end{table}
}

\eat{Building on the above discussion, we}This paper proposes \oursys, an in-memory system for dynamic graph processing that employs an interleaved execution mode combining coroutines and software prefetching.
\oursys focuses on graph storage and coroutine-based prefetch co-design, proposing a novel dynamic graph data structure friendly to prefetching techniques.
Besides, \oursys comprises various task allocating, prefetching, and scheduling strategies to mitigate the overhead of cache misses, thereby delivering robust performance for different graph processing tasks across various hardware environments. %\gsf{see comment}%只介绍基本的idea就行吧。设计了castcoco，有啥优点，next present the details.

As shown in \autoref{fig:gastcoco}, \oursys is primarily composed of three layers: storage layer, execution layer, and adaptation layer.
Given a graph processing task (\ie graph computation or graph update), user execution begins by calling \oursys API, as shown in \autoref{tab:api}.
\oursys API activates the dynamic graph processing engine, at which point the system configuration probe in the adaptation layer detects the hardware environment to configure the system parameters of \oursys, \eg the numbers of coroutines per thread and hybrid prefetching strategy.
To process tasks using software prefetching, graph data should be partitioned into several parts and allocated to a coroutine within a coroutine pool.
All coroutines collaborate according to the scheduling strategy to produce the final results.
The execution strategy tuner tailors the partition and scheduling strategy to suit the specific task to achieve better performance.
The task allocator partitions the graph data according to the partition strategy and constructs the coroutine pool, while the coroutine scheduler schedules coroutines according to the scheduling strategy.
% Initially, \oursys adjusts prefetching strategy and coroutines numbers through pre-configured programs for the current hardware environment in the adaptation layer, preparing for the subsequent execution.
% Next, in the execution layer, \oursys tailors the scheduling and computation to suit the specific task to achieve better performance.
Finally, the executor performs all read and write operations on the storage layer.

% \noindent
% \textbf{\oursys API.} \zyf{API Layer} \zyf{the apis should be very carefully revisited, it is very important. Now, it seems too ....}
% We offer a set of APIs for dynamic graph processing.
% For graph computation, users can utilize the APIs in conjunction with vertex and edge access operations (summarized in \autoref{subsec:acc_pattern}) to complete existing graph computation tasks on \oursys.
% For graph updates, users can utilize the APIs to load graph files, update, and update equipped with software prefetching via coroutine in batch.

\noindent
\textbf{Storage Layer (\hyperref[sec:storage-layer]{§4}).} %Prefetch-Aware Data Structure
This layer is our specially designed data structure \ourstorage, which supports efficient graph computation and frequent graph updates.
%Our prefetch-aware data structure stores the entire graph by storing the neighboring edges for each vertex. %这句话貌似没用
%\gsf{To support efficient queries and intersections, in \ourstorage, each vertex's edges are maintained in an ordered fashion.} %to support efficient queries and intersections.
%\gsf{In addition, the design of \ourstorage tends to favor hardware prefetching, since it is more efficient and has no additional overhead compared to software prefetching.}
The design of \ourstorage is made to align as closely as possible with hardware prefetching mechanisms (as discussed in \autoref{subsec:cache_miss_and_hard_pre}) for reducing cache misses. 
%The design of \ourstorage is made to align as closely as possible with hardware prefetching mechanisms discussed in \autoref{subsec:cache_miss_and_hard_pre}, such that the hardware prefetching works and achieves better performance during computation.

\begin{table}[t!]
    \caption{\oursys APIs.}
    \label{tab:api}
    % \small
    \footnotesize
    \setlength{\tabcolsep}{1.7pt}
    \begin{tabular}[t]{lll}
        \toprule
        Name & Parameters & Description \\
        \midrule
        \textbf{LoadGraph} & (file\_path) &  Load graph data files. \\
        \textbf{UpdateVertex} & (obj\_v, update\_mode) & Add, delete, and modify vertices. \\
        \textbf{UpdateEdge} & (obj\_e, update\_mode) & Add, delete, and modify edges. \\
        \textbf{BatchUpdate} & (update\_content) & Handle batch update tasks. \\
        \midrule
        \textbf{ProcessVertex} & (f, active) & \eat{Implementing} Process vertices task.\\
        \textbf{ProcessEdge} & (dense\_f, sparse\_f, active) & \eat{Implementing} Process edges task. \\
    \bottomrule
    \end{tabular}
    \vspace{-0.1in}
\end{table}

%\todo{lhf-flag-for ta}
\noindent
\textbf{Execution Layer (\hyperref[sec:exe-layer]{§5}).}
The execution layer includes the task allocator which breaks down a task into multiple subtasks and allocates them to coroutines, the coroutine scheduler which controls the interleaving execution and prefetching, and the executor which executes the computation tasks according to the scheduling strategy.
%The components in the dynamic graph processing engine include task allocator, coroutine scheduler, and executor.
% They help ensure that the performance of \oursys is optimized for the current runtime environment and the specific graph processing task at hand.
%Depending on the specific graph processing task, we conduct a detailed partition and scheduling strategy based on analyzing data access operations in the task allocator.
%Subsequently, it allocates subtasks and constructs coroutines according to the partition strategy. Finally, the executor and the coroutine scheduler execute the graph processing tasks following the established scheduling strategy.

%\todo{lhf-flag for est}
\noindent
\textbf{Adaptation Layer (\hyperref[sec:adapt-layer]{§6}).}
The adaptation layer includes two components: the system configuration probe and the execution strategy tuner.
They utilize prefabricated programs to probe the optimal coroutine parameters (\eg number of coroutines per thread) and the most suitable execution and prefetching strategy for the current tasks and runtime environment.
%Of course, manual setting of fixed parameters by users is also allowed.

% \gsf{see comment}%是不是应该拿一个任务来说事，然后从上到下介绍整个流程。

\section{Prefetch-Aware Structure \ourstorage}
%This section presents the proposed prefetch-aware structure \ourstorage.% and then details the edge storage and global traversal chain design.

\label{sec:storage-layer}
% In this section, we introduce \todo{...} in dynamic graph processing scenarios.

\subsection{Overview}

For dynamic graph processing tasks, we have the following expectations for an ideal data structure.
It should: 
(1) maintain cache locality as possible to produce efficient graph computation;
(2) be easy to dynamically adjust for frequent graph updates.
% (3) be more conducive to hardware prefetchers to predict\gsf{conducive is suitable for here? I am not sure};
% (4) be more friendly to the execution of strategy formed by coroutines and software prefetching \gsf{not understand}. \gsf{see comment}%我理解不就是两个吗？一个是计算，一个是更新，计算的时候3 4是为了计算？
%(we will elaborate on this item in \todo{sec} and not expand upon it in this section).
% This section focuses on (1)-(3) and (4) will be discussed in \autoref{subsec:LoadBalance} \gsf{why move to sec 5? not friendly to reader}.
To enhance the performance from both perspectives, a basic idea\eat{ of our data structure design} is to tailor as closely as possible to fit the hardware prefetching mechanisms and prepare for the implementation of software prefetching\eat{ via coroutines}.
% We omit the details of deleting operations because the deleted vertices and edges can be marked by flags and then handled by garbage collection (GC) in the background.
% to achieve high update throughput, we employ deletion flags for vertex or edge deletion.
% The physical deletion and memory free-up are handled by garbage collection (GC) in the background.

\begin{figure}[t]
  \centering
  \includegraphics[width=0.85\linewidth]{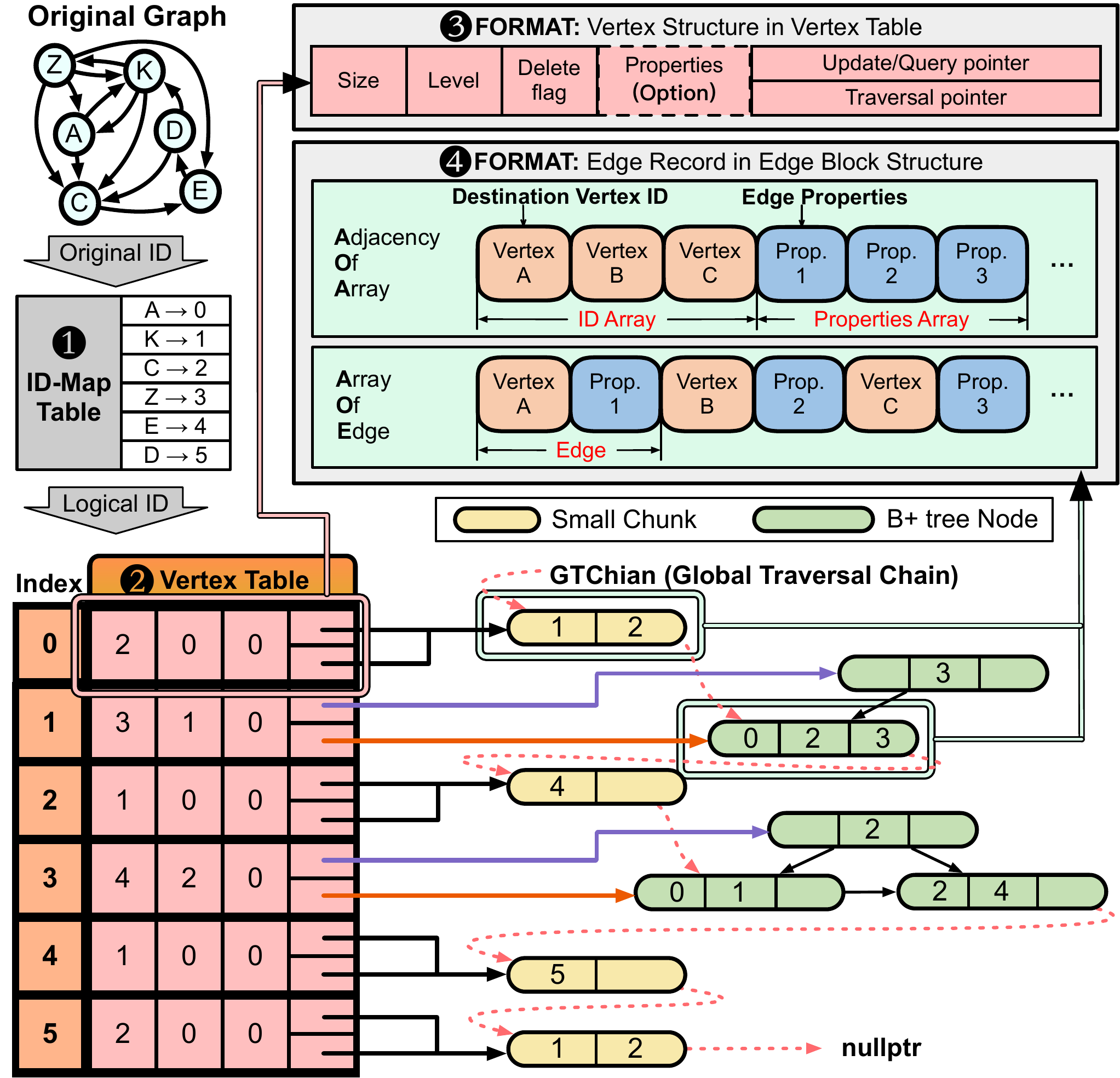}
  \vspace{-0.1in}
  \caption{Prefetch-aware structure \ourstorage. (The capacities of small chunks and B+ tree nodes are set as 2 and 3 edges.)}
  \label{fig:cblist}
  \vspace{-0.1in}
\end{figure}

\autoref{fig:cblist} presents the proposed data structure \ourstorage and we elaborate on its details as follows. 
For \textit{storing vertices}, we employ an ID-map table and a vertex table (array of vertex structures), as shown in \autoref{fig:cblist} {\blackcircleone} and {\blackcircletwo}, to map the vertex's original ID \eat{(any data type) }to the logical ID (unique number identifier used in graph computation).
% The ID-map table is used to map the vertex's original ID to the logical ID. In the vertex table, the logical ID is used as an index to store the vertex records.
A record contains multiple fields as presented in \autoref{fig:cblist} \blackcirclethree, which includes two pointers to the neighbors (\ie traversal pointer and update/query pointer), size (number of adjacent edges), delete flag, and the level.
Here, we use a level variable to distinguish the structure of the vertex's neighborhood. 
A value $0$ represents using small chunks for edge storage, while other numbers indicate using B+ trees and the value represents the number of leaf nodes in the tree.

% Next, let's discuss the advantages of Vertex Storage design.
% From the perspective of requirements, logical IDs ensure that adding vertices to the Vertex Table can almost be regarded as an append operation, eliminating the data movement overhead caused by an insert operation. 
% This is because the ID-map Table can map the ID of each newly added vertex to the current maximum logical ID plus one. 
% Correspondingly, when adding to the Vertex Table, it becomes an append operation to the end, rather than an insert to the middle.
% \PrOneV completes in $O(1)$ by using logical IDs as an array index.
% Besides, the design of the Vertex Table for contiguous memory conforms to the scanning of vertices and makes \PrAllVs efficient.
% Under the data organization of arrays, we are able to design coroutine schedulers with more balanced workloads and the data structures have more flexible extension ability to distributed systems. 
% When dividing sub-tasks for coroutines and considering future work to be extended to distributed systems, partitioning based on an array is much faster compared to partitioning based on a tree structure.
% When the number of coroutines, threads or distributed machines dynamically changes (possibly by expanding or reducing), to split an array will not be a time-consuming operation. \zyf{comment}

% \textbf{Edge Storage Design.}
For \textit{storing edges}, we design the update-read balanced edge storage with hierarchical structures, as shown in the lower right of \autoref{fig:cblist}.
The details will be introduced in \autoref{subsec:urbalanced}.
The edges of each edge block structure are organized as shown in  \autoref{fig:cblist} \blackcirclefour.
Two pointers, namely the traversal pointer and update/query pointer, are utilized to link each edge storage and the vertex table. 
% For a high-degree vertex, the traversal pointer (in purple arrows) and update/query pointer (in red arrows) point to the leftmost leaf node and the root node of B+ tree, respectively.
% For a low-degree vertex, both pointers (the black arrows from \autoref{fig:cblist} \blackcircletwo) point to the start of small chunks. \gsf{Why two points?}
We provide two edge property storage modes: AOE (Array Of Edge) and AOA (Adjacency Of Array), as depicted in \autoref{fig:cblist} \blackcirclefour.
% \todo{lhf: add explain}
% AOA is an efficient edge storage if edge properties are not needed (\eg algorithms like PageRank, BFS, DFS, etc.).
% AOE may perform better if most or all edge properties are needed during the computation.

Finally, we also design a prefetch-friendly global traversal chain to fully take advantage of hardware prefetching for graph traversal, the details of which can be found in \autoref{subsec:GTChain}. %\zyf{most of the description can be included in the figure or in the fig captain.}

% \tao{see comments}
% % 介绍vertex和edge property对后面有帮助吗？这两段感觉可有可无，可以缩短，两句话介绍即可
% % \textbf{Vertex Properties.} \gsf{not your focus, considering delete it}
% The properties of vertices are an optional feature within the vertex structure, not explicitly represented in \ourstorage figure. \zyf{which figure} 
% Whether they are stored can be freely chosen based on requirements. 
% For instance, if the storage space for vertex properties is excessively large, one might consider allocating a separate array to store them and using logical IDs as indices for correspondence.

% \textbf{Edge Properties.}

\subsection{Update-Read Balanced Edge Storage}
\label{subsec:urbalanced}
% \subsection{Edge Storage Design}
% 只介绍advantages很奇怪，至少应该介绍一下edge storage design，或者将edge storage design的主要内容移到这一部分。
%As an 
Array-like structure, \eg CSR, suffers from poor graph update performance due to data movements. % during insertions.
However, we found that %the cost of data movements is related to the scale of the data. 
for small data volumes, the data movement cost is negligible since only a few data moves. Meanwhile, the continuous memory Array takes full advantage of CPU cache performance to significantly improve graph computing efficiency. For larger data volumes, incorporating non-contiguous memory designs should be considered to mitigate the cost of data movements.
This means we can construct a hierarchical structure to achieve the update-read balance. In real-world graph scenarios, the number of edges per vertex varies, thus a refined design for edge storage is necessary.

\textbf{Basic Idea.}
We use small chunks (\ie yellow capsule-shaped structures in \autoref{fig:cblist}) to store %for 
low-degree vertices, and employ B+ trees~\cite{DBLP:journals/csur/Comer79} (\ie green capsule-shaped structures in \autoref{fig:cblist}) to store %for 
high-degree vertices.
The capacities of small chunks and B+ tree nodes are 
set as integer multiples (usually $1$-$4$) of the cache line size.
Specifically, we set the multiples as $4$ and the destination nodes are inserted into the chunk with the update of the graph.
When the data exceeds the capacities of small chunks, we reorganize it in a B+ tree with tree node size also multiples of cache line size.
For better readability, we set the capacities of small chunks and B+ tree nodes to 2 and 3 edges, respectively, in \autoref{fig:cblist}.
While in practice, we make the size of the chunks and tree nodes aligned with the cache line, such that the structure can provide better cache locality and enhance cache utilization for graph computation.
% B+ tree nodes are fixed at $4$ times the cache line size.
% This also serves as the boundary for the conversion between the two structures.
% When data volume exceeds this boundary, we do not expand the small chunk. Instead, we treat it as a B+ tree node.
% Additionally, aligning the size with its cache line size can provide better cache locality and enhance cache utilization for
% graph computation.

\begin{figure}[t]
  \centering
  \includegraphics[width=0.9\linewidth]{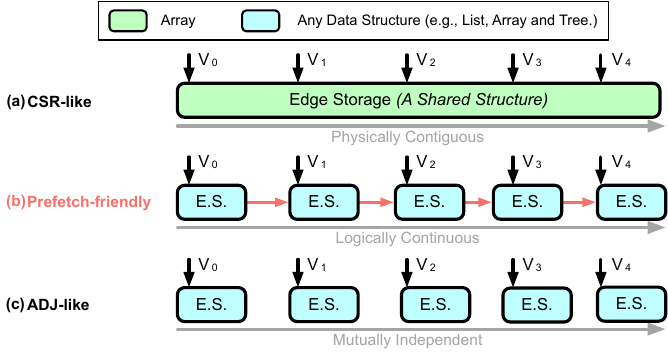}
  \vspace{-0.1in}
  \caption{(a) CSR-like physically contiguous store; (b) prefetch-friendly logically contiguous store via pointers; (c) ADJ-like mutually independent store (E.S.: Edge Storage).}
  \label{fig:ADJ-and-CSR}
  \vspace{-0.1in}
\end{figure}

We enumerate the advantages of our Edge Storage design as follows.
Firstly, we use independent edge structures (refer to ADJ-like structures in \autoref{fig:ADJ-and-CSR}) for each vertex to store its neighborhood to promote parallel updates of edges and facilitate different implementations like chunks and B+ trees for different nodes.
In contrast, in CSR-like structures (\eg PCSR~\cite{DBLP:conf/hpec/WheatmanX18} and Teseo~\cite{de2021teseo}), nodes are stored in identical and shared edge structure (refer to the CSR-like structures in \autoref{fig:ADJ-and-CSR}), and to update neighbors of one vertex may affect the vertex records and memory locations of other edges.
Secondly, \PrAllNeis accessed from Traversal pointer in a B+ tree can be considered as sequential data access on a linked list, which can effectively utilize hardware prefetching as discussed in \autoref{subsec:cache_miss_and_hard_pre}.
Thirdly, \PrOneNei accessed from update/query pointer completes in $O(log(n))$.

\subsection{Prefecth-Friendly Global Traversal Chain}
\label{subsec:GTChain}

% \textbf{Global Graph Design.}
In graph computation, a significant portion of tasks (\eg PageRank~\cite{page1998winograd}) require all vertices and edges in the graph to be involved in the computation. 
Their data access patterns can be represented as a combination of \PrAllVs and \PrAllNeis.
This combination equates to sequential data access over the entire graph.

From a global graph perspective, although CSR-like structures with shared edge structures impact update performance, they store all graph data in contiguous memory as shown in \autoref{fig:ADJ-and-CSR}. 
In contrast, ADJ-like structures, despite their independently stored vertex neighborhoods being update-friendly, result in complete fragmentation of the entire graph data in memory.

\textbf{Basic Idea.} We consider a transitional form:
based on the data structure, we connect the neighborhoods of vertices with adjacent logical IDs to form a linked list (the chain formed by red dashed lines in \autoref{fig:cblist}) that represents the entire graph named \textit{Global Traversal Chain} (GTChain).
Consequently, it can naturally leverage hardware prefetching to significantly improve the performance during sequential access (composed of \PrAllVs and \PrAllNeis) to entire graph data.
Additionally, it provides assistance in load balancing for coroutines, as will be detailed in \autoref{subsec:LoadBalance}.

We have also considered the overhead of forming GTChain.
One of the reasons we chose B+ trees is the ease of forming GTChain.
Each B+ tree inherently contains a traversal chain formed by leaf nodes.
Its implementation for GTChain simply involves using pointers to link the leaf node chain to the global chain. 
However, most tree structures (\eg balanced binary trees, B-trees, and red-black trees) require manual implementation of a traversal chain and the additional traversal chain is challenging to form as a singly linked list (each node has only one outgoing pointer) to support the jump-pointer mechanism of hardware prefetching.

\section{Interleaved execution with coroutine}
\label{sec:exe-layer}

% This section...
%first presents an analysis of the workload and the memory access pattern in dynamic graph processing and then demonstrates the necessity of discontinuous memory in the data structure. Second, we discovered optimizations in modern CPUs for linked list traversal. We leverage the hardware prefetching mechanism to accelerate the workload in Graph. Third, we further observed that the hardware prefetching mechanism may fail in certain scenarios of the workload in Graph. To address this issue, we introduced a combination of coroutines + software prefetching techniques to compensate for the performance impact caused by the failure of hardware prefetching.

\subsection{Coroutine with Software Prefetching}
\label{subsec:interleave-exe}

% \gsf{see comment}%gsf: 整个subsec感觉跟图也没太大关系啊，sequential和interleaved execution不用graph也能说，包括后边的coroutine也跟graph没啥太大关系。

\begin{figure}[tbp]
% \vspace{-0.15in}
  \centering
  \includegraphics[width=\linewidth]{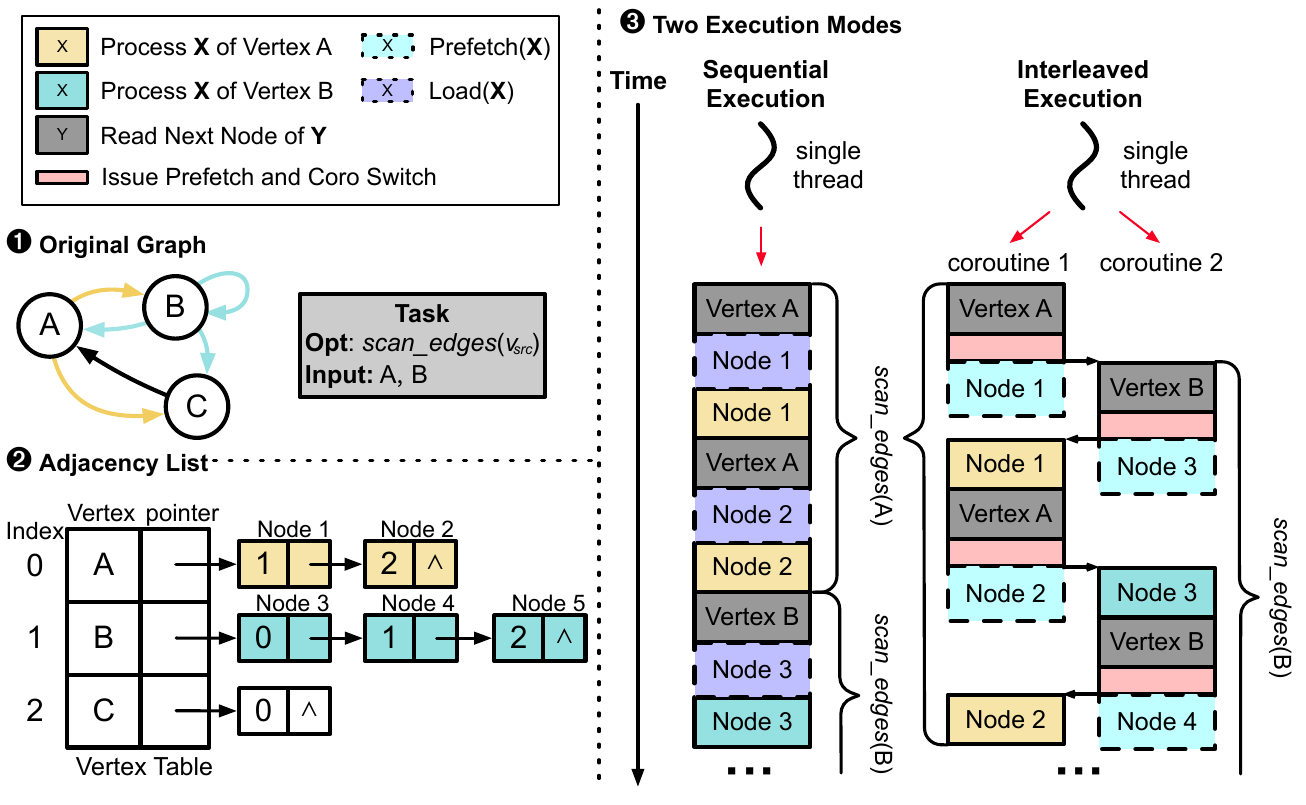}
  \vspace{-0.2in}
  \caption{Sequential execution vs. interleaved execution.}
  \label{fig:overlap_interleave}
  \vspace{-0.1in}
\end{figure}

\etitle{Sequential execution and interleaved execution}. 
This paper focuses on reducing the cache miss overhead based on the interleaving execution mode~\cite{DBLP:journals/pvldb/PsaropoulosLMA17}.
\autoref{fig:overlap_interleave} illustrates the difference between sequential execution and interleaving execution of traversal on a graph in \autoref{fig:overlap_interleave} \blackcircleone. 
% Since we have not yet introduced coroutine task allocation and construction \gsf{move: Since XXX construction}, 
\eat{To improve the understanding of the execution process, w}
We first demonstrate an example of \PrAllNeis on AL (adjacency list) in \autoref{fig:overlap_interleave} {\blackcircletwo} for clarity:
In the sequential execution mode, when the CPU accesses Node 1, the first neighbor of vertex A, a cache miss is likely to occur in the non-contiguous memory linked by pointers, making the CPU stalled, \ie waiting until the data is fetched\eat{ while waiting for the data}. 
Such stalling~\cite{he2020corobase} occurs whenever cache misses happen due to non-contiguous memory access and pointer chasing.
Alternatively, in interleaved execution mode, the CPU will issue a prefetch for Node 1 and switch to another task (\ie accessing Node 3, the first neighbor of vertex B).
\eat{As a result, for sequential execution, whenever the cache misses occur for accessing non-contiguous memory and pointer chasing, the CPU will experience data stalls~\cite{he2020corobase} and must wait until the data is fetched in the cache.}
Each coroutine handles a linked list and is in charge of computing the corresponding task.
The algorithm starts from an arbitrary computation task.
It switches to another task whenever it encounters a software prefetching instruction to fetch data (the instruction is set before accessing the next node in the linked list).
When switching back to the previous task, the CPU does not need to wait for the needed data as they have been loaded into the cache.
Thus, the process of fetching data and computations can be overlapped and the time of cache misses due to pointer chasing can be avoided.
Note that switching tasks also causes time costs~\cite{DBLP:journals/pvldb/PsaropoulosLMA17}.
We choose C++20 stackless coroutines and design scheduling strategies, such that the switching time is minimal, \eat{less than the time of occurring data stalls }thereby achieving performance enhancements.
% our workload desires neighbor blocks A and B of vertex 0, as well as neighbor blocks C, D, and E of vertex 1. 
% If executed sequentially, the CPU often experiences data stalls~\cite{he2020corobase} due to cache misses caused by non-contiguous memory access and pointer chasing, which means the CPU has to wait.
% However, in the interleaved execution mode, when encountering a pointer, the CPU issues a data-fetching instruction and switches to another computational task instead of waiting for the data. 

\etitle{Coroutine in Graph Computation.}
Next, we present the coroutine pool constructor and scheduler in graph computation.
As we previously mentioned, C++20 stackless coroutines are a special kind of function and it requires manual intervention to schedule these functions.
\ding{172}In \autoref{alg:simpleScl} (\autoref{alg1-constructor:start} - \autoref{alg1-constructor:end}), we present the basic logic of the coroutine pool constructor.
The time complexity of the constructor is $O(m)$, where $m$ is the number of coroutines in the coroutine pool.
The coroutine pool relies on the for loop (\autoref{alg1-constructor:for}) to initialize each coroutine as \textit{Func}.
\textit{Func} defines the graph access operations (as proposed in \autoref{subsec:cache_miss_and_hard_pre}) which will be detailed in \autoref{alg:get_e}.
% \textit{Func} represents a function containing the basic logic of the coroutine.
% To utilize coroutines in graph computation, we model graph access operations (as proposed in \autoref{subsec:cache_miss_and_hard_pre}) as coroutines (\textit{Func}), which will be detailed in \autoref{alg:get_e}.
\ding{173}In \autoref{alg:simpleScl} (\autoref{alg1-sche:start} - \autoref{alg1-sche:end}), we present a polling scheduler implementation.
The overhead generated by the scheduler itself is also a cost of using coroutines.
Therefore, we design it as simply as possible to reduce the cost of using coroutines.
Its core logic includes checking if the coroutine pool has finished through a while loop (\autoref{alg1-sche:while}) and polling each coroutine via a for loop (\autoref{alg1-sche:for}).
If a coroutine is not completed, the polling scheduler resumes execution from the suspension point (\autoref{alg:simpleScl:scheduler-resume}).
If a coroutine is completed, the polling scheduler will destroy it (\autoref{alg1-sche:des}). 
Assuming the number of suspensions for the $i$-th coroutine in the polling scheduler is $k_{i}$, the time complexity for the entire polling scheduler is $O(\sum_{i=1}^{m}  k_{i})$.
If the tasks are evenly distributed and each coroutine is suspended approximately $k$ times, the overall time complexity is $O(mk)$.
%\tao{two sketchy. see comments}
% 关于算法的描述过于简单，单纯通过伪代码很难理解。至少将描述与伪代码中的while等关键字或者某几行匹配
% However, the performance of such a scheduler with simple logic heavily depends on whether the division of coroutine sub-tasks is even and balanced. 
% We further propose a fine-grained scheduler in \autoref{subsec:LoadBalance}\lhf{need relocate} to achieve a better performance.
% \ding{174}
% Since vertex access operations occur in the vertex table (Array of Vertex Structures in \ourstorage), there is no need for software prefetching because of the good cache locality of \ourstorage.
% Therefore, we do not directly construct them using coroutines.
% However, it is noteworthy that edge access operations are constructed using coroutines, and since we enter edge access operations through vertex access operations, it implies that vertex access operations indirectly involve coroutines.
% We provide the logic of vertex access operations in \autoref{alg:get_v}.
% \GV, \GVc and \FV correspond to vertex access operations \PrAllVs, \PrAllVsCond, \PrOneV.
% \lhf{no details now, may delete, need discuss with prof.zhang }
% \tao{similarly to Algorithm 1, almost no details of Algorithms 2}
\ding{174}We detail the logic of edge access operations in \autoref{alg:get_e}. 
\GEv and \GEc correspond to the \PrAllNeis operation for accessing edges and their \eat{difference between \GEv and \texttt{GetNeighbors}
\\
\texttt{($chain$)}\xspace and their }usage and difference will be elaborated in \autoref{subsec:coro_represent}. 
The time complexities of the two operations are $O(D)$ and $O(E_{chain})$ respectively, where $D$ is the degree of the vertex, and $E_{chain}$ is the number of edges in the chain.
\texttt{FindNeighbor}\texttt{($edge$)} corresponds to \PrOneNei.
Based on \ourstorage, regardless of whether using a binary search on small chunks or searching within a B+ tree, the time complexity for \texttt{FindNeighbor}\texttt{($edge$)} is $O(\log D)$.
The user can specify the search algorithm either on small chunks or B+ tree leaf nodes (refer to \autoref{alg2-search1}, \autoref{alg2-search2}).
Here, all edge access operations are constructed using coroutines and
\texttt{prefetch($ptr$)} denotes prefetching the content pointed to by the pointer in \autoref{alg:ge1p}, \autoref{alg:ge2p}, and \autoref{alg:ge3p}.
This is in the form of software prefetching.
In \autoref{alg:ge:suspend-1}, \autoref{alg:ge:suspend-2}, and \autoref{alg:ge:suspend-3}, \emph{co\_await} is a keyword in C++20 and its invocation of \textbf{suspend\_always} means that this coroutine actively entering an always suspended state.
Coroutines in the suspended state will be resumed by the scheduler (\autoref{alg:simpleScl:scheduler-resume} in \autoref{alg:simpleScl}).

\begin{algorithm}[t]
\DontPrintSemicolon
\caption{Constructor \& Polling Scheduler}
\label{alg:simpleScl}

\SetKwFunction{FScheduler}{Scheduler}
\SetKwFunction{Fconstructor}{Constructor}
\SetKwFunction{Fappend}{append}
\SetKwFunction{Fdone}{done}
\SetKwFunction{Ffinish}{finish}
\SetKwFunction{Fresume}{resume}
\SetKwFunction{Fdestory}{destroy}
\SetKwProg{Fn}{Function}{:}{}
\SetKw{KwTo}{to}
\Fn{\Fconstructor{$coroutine\_num$, $Func$}}{ \label{alg1-constructor:start}
    \For{\( i \gets 1 \) \KwTo  $coroutine\_num$}{ \label{alg1-constructor:for}
        $coroutine\_pool.$\Fappend{$Func$}\;
    }\label{alg1-constructor:end}
}
\;
\Fn{\FScheduler{$coroutine\_pool$}}{\label{alg1-sche:start}
    \While{not $coroutine\_pool.$\Ffinish{}}{ \label{alg1-sche:while}
        \For{\( i \gets 1 \) \KwTo  $coroutine\_pool.size$}{ \label{alg1-sche:for}
            \If{not $coroutine\_pool[i].$\Fdone{}}{
                $coroutine\_pool[i].$\Fresume{}\; \label{alg:simpleScl:scheduler-resume}
            } 
            \Else{
                $coroutine\_pool[i].$\Fdestory{}\;
                \label{alg1-sche:des}
            }
        }
    }\label{alg1-sche:end}
}
\vspace{-0.05in}
\end{algorithm}

\etitle{Coroutines in Graph Update.}
\oursys primarily aims to enhance performance in the context of \emph{batched edge updates} by utilizing coroutines. 
For other updates, we have the following discussions.
\eat{Other updates can be implemented efficiently with straightforward implementations.}
(1) \eat{\oursys supports two categories of graph updates, including individual updates and batch updates.}
For individual updates, as previously mentioned, there is no overlapping execution with other tasks to diminish the data loading overhead.
Therefore, it cannot form an interleaved execution mode and leverage coroutines to optimize performance.
However, \oursys can still rely on the inherent efficiency of \ourstorage for handling individual updates.
(2) Furthermore, graph updates can be divided into vertex updates and edge updates.
For vertex updates, the vertex table in \ourstorage makes it efficient to query a vertex by the operation \PrOneV.
The vertex insertion operation in the vertex table can be achieved by an append operation because we can align the new vertex to the maximum logical ID based on the ID-map table. 
In this way, we do not use coroutines for vertex updates in the vertex table either.
% To sum up, when applying coroutines to graph updates, we only consider batch edge updates.

Specifically, in \oursys, no matter whether an update is an insertion, deletion, or modification, the process is divided into two steps: \textbf{locating} and \textbf{operating}.
We first locate the position to be updated through a query operation, and then proceed with the specific update operation.
Thus, we formulate the query operation \PrOneNei as a coroutine, as shown in \autoref{alg:get_e} (\autoref{alg2:FN}).
During the queries in \ourstorage, whenever we meet a pointer for data loading, we issue a software prefetch (\autoref{alg:ge3p}) and switch back later to hide the overhead of loading data into the cache.

\begin{algorithm}[!t]
\DontPrintSemicolon
\caption{Edge Access Operation}
\label{alg:get_e}

\SetKwFunction{FGetNeighbors}{GetNeighbors}
\SetKwFunction{FGetVs}{GetVertices}
\SetKwFunction{FFindV}{FindVertex}
\SetKwFunction{Fcompute}{compute}
\SetKwFunction{Fprefetch}{prefetch}
\SetKwFunction{Fcompute}{compute}
\SetKwFunction{FFindNeighbor}{FindNeighbor}
\SetKwFunction{Fsearch}{search}
\SetKwFunction{Flocate}{locate}
\SetKwProg{Fn}{Function}{:}{}
\SetKwFor{For}{for}{do}{endfor}
\SetKw{KwTo}{to}
\SetKw{KwRet}{Return}
\SetKwIF{If}{ElseIf}{Else}{if}{:}{else if}{else}{endif}

% // \PrAllVs \;
% \Fn{\FGetVs{}}
% // \PrAllVsCond \;
% \Fn{\FGetVs{$cond$}}
% // \PrOneV \;
% \Fn{\FFindV{$vertex$}}

// \PrAllNeis \;
\Fn{\FGetNeighbors{$vertex$}}{
    $n \gets 1$\;
    \If{$VertexTable[vertex].level \neq 0$}{
        $n \gets VertexTable[vertex].level$
    }
    $ptr \gets VertexTable[vertex].Traversal\_pointer$\;
    \For{\( i \gets 0 \) \KwTo \( n \)}{
        \Fprefetch{$ptr$}\; \label{alg:ge1p}
        co\_await \textbf{suspend\_always}\; \label{alg:ge:suspend-1}
        \Fcompute{$ptr$}\;
        $ptr \gets ptr.next$\;
    }
}
// \PrAllNeis \;
\Fn{\FGetNeighbors{$chain$}}{
    $ptr \gets  chain.start$\;
    \While{$ptr \neq chain.end$}{
        \Fprefetch{$ptr$}\; \label{alg:ge2p}
        co\_await \textbf{suspend\_always}\; \label{alg:ge:suspend-2}
        \Fcompute{$ptr$}\;
        $ptr \gets ptr.next$\;
    }
}
// \PrOneNei \;
\Fn{\FFindNeighbor{$edge$}}{ \label{alg2:FN}
    $SmallChunk\_level \gets 0$\;
    $ptr \gets VertexTable[edge.src].Query\_pointer$\;
    \If{$VertexTable[edge.src].level = SmallChunk\_level$}{
        \KwRet \Fsearch{$ptr$, $edge.dst$}\; \label{alg2-search1}
    }
    \Else{
        \While{$ptr \neq nullptr$}{
            \Fprefetch{$ptr$}\; \label{alg:ge3p}
            co\_await \textbf{suspend\_always}\; \label{alg:ge:suspend-3}
            \If{$ptr.Type = LeafNode$}{
                \KwRet \Fsearch{$ptr$, $edge.dst$}\; \label{alg2-search2}
            }
            \Else{
                $ptr \gets$ \Flocate{$ptr$, $edge.dst$}\;
            }
        }
    }
    \KwRet $False$\;
}
\vspace{-0.05in}
\end{algorithm}

\subsection{Load Balancing of Coroutines}
\label{subsec:LoadBalance}

\begin{figure}[!t]
% \vspace{-0.1in}
  \centering
\includegraphics[width=\linewidth]{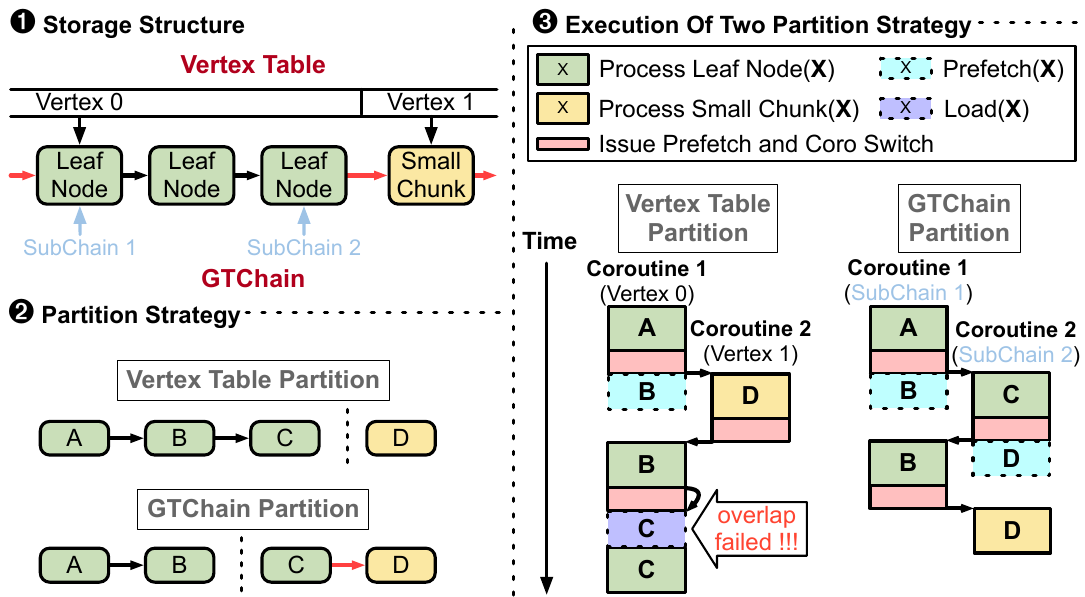}
\vspace{-0.2in}
  \caption{Load balancing of coroutines. }
  \label{fig:loadbalance}
  \vspace{-0.2in}
\end{figure}

As we have described in the example of interleaving execution mode in \autoref{fig:overlap_interleave}, each coroutine in a graph processing task represents a unit of concurrent execution that processes the subgraph from the overall task.
We can reduce the overhead of cache misses by overlapping the time of fetching data through the switching of different coroutines.
However, switching coroutines also causes an overhead.
Suppose a coroutine requires much more switch times than other coroutines.
Under the scheduling by polling scheduler (\autoref{alg1-sche:start} in \autoref{alg:simpleScl}), it will occupy the most time of the algorithm's execution, with other coroutines having no overlapping with it.
In such case, it will lead to an unbalanced interleaving execution mode, as illustrated in ``Vertex Table Partition'' in \autoref{fig:loadbalance} \blackcirclethree.

To this end, two solutions can solve the issues. The first is to avoid the unbalanced partition by ensuring that each coroutine has approximately the same switch times, and another is to optimize the polling scheduler to avoid the repeated suspension and resumption of the remaining single coroutine after all other coroutines have been destroyed.
Correspondingly, we propose two techniques to address the issue: a graph partition strategy to ensure partition evenness and a trimmed-polling scheduler with fine-grained checkpoints.
``Vertex Table Partition'' is the basic graph partition strategy by dividing the vertex table into continuous chunks, which could maintain the natural locality of the graph data ~\cite{DBLP:conf/osdi/ZhuCZM16}.
% More specifically, there are two approaches.
% \ding{172}The first is \textit{average vertex number partition} with lower division costs.
% Suppose there are $N$ coroutines and $V$ vertices, then each coroutine will process the edges of $V/N$ vertices (a continuous chunk divided from VertexTable).
% For most of the graph data, this approach can result in a more balanced partition. 
% However, in some heavily skewed graph data, there will be only one or few super vertices with large degrees, while most vertices have few neighbors.
% This often results in the situation depicted in \autoref{fig:loadbalance}.
% \ding{173}The second is \textit{average edge block number partition}, aiming to balance the number of edge structure blocks (small chunks or leaf nodes) processed by each coroutine.
% To this end, we maintain an additional prefix sum array storing the number of edge blocks.
% Assuming there are $N$ coroutines and \ourstorage contains $X$ edge structure blocks, each coroutine will target $X/N$ edge blocks for partition. 
% We search from the beginning of the prefix sum array and assign the continuous chunks containing the closest to $X/N$ edge blocks to the coroutines in sequence.
 This approach may also lead to unbalance in some extreme cases, as partitioning on the vertex table means that the smallest granularity of the partition is the edge block of a single vertex.
For example, if a vertex is a super vertex (\eg having $95\%$ of the edges), the approach may fail.
% So we need to consider an alternative partition strategy instead of using Vertex Table Partition.

\textbf{A Fine-grained Graph Partition Strategy}, on the other hand, controls the balance by partitioning GTChain into continuous sub-chains, as shown in \autoref{fig:loadbalance} \blackcircletwo.
Suppose there are $N$ coroutines and \ourstorage contains $X$ edge structure blocks, then each coroutine will process a sub-chain (cut from GTChain) containing $X/N$ edge blocks.
While this strategy achieves nearly perfect balance, it requires the participation of all the vertices and edges in the entire graph computation, just like the conditions for using GTChain.
However, even though we evenly divide the data before computation, if certain edge blocks are not involved in the computation, it still cannot form a balanced interleaving execution mode at run time.
As a result, ``GTChain Partition'' can only be used for \PrAllVs with \PrAllNeis.
In summary, we design and employ two graph partition strategies in \oursys, and their specific usage will be detailed in \autoref{subsec:coro_represent}.

\textbf{A Trimmed-Polling Scheduler.}
Compared with the polling scheduler, a more proper approach would be to pass the number of uncompleted coroutines to the resumed coroutine such that it can decide when to switch. 
If the coroutine is the only one that remains working, it would not issue a suspension operator, as shown in \autoref{alg:complexScl}. 
On the contrary, this may introduce extra overhead because of the additional checks in practice. 
% In the time-critical execution process, it is more desirable to avoid such an approach if possible.

\section{Adaptation Layer}
\label{sec:adapt-layer}

\subsection{Execution Strategy Tuner}
\label{subsec:coro_represent}

\begin{figure}[t]
  \centering
  \includegraphics[width=0.9\linewidth]{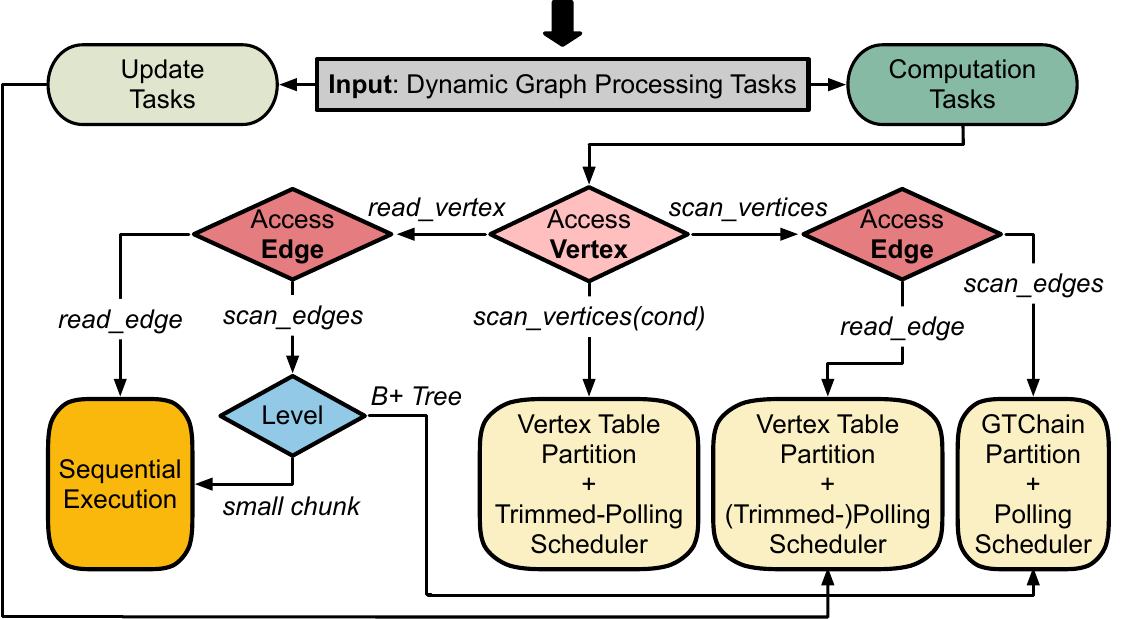}
  \vspace{-0.1in}
  \caption{The flow of execution strategy tuner.}
  \label{fig:CoroAllocator}
\vspace{-0.1in}
\end{figure}

In this subsection, we explain how the execution strategy tuner tailors partition and scheduling strategies based on graph processing tasks and how the task allocator allocates tasks to coroutines.
It is noteworthy that modeling each task as a coroutine and creating the interleaved execution mode among multiple tasks is feasible and easy.
Thus, we focus on building interleaved execution patterns among multiple coroutines within a single task in this subsection.
We present the logic of the execution strategy tuner in \autoref{fig:CoroAllocator} for a better understanding.

\begin{algorithm}[!t]
% \vspace{-0.1in}
\DontPrintSemicolon
\caption{Trimmed-Polling Scheduler}
\label{alg:complexScl}

\SetKwFunction{FScheduler}{Scheduler}
\SetKwFunction{FCoro}{Coroutine\_Func}
\SetKwFunction{Ffinish}{finish}
\SetKwFunction{Fresume}{resume}
\SetKwFunction{Fdestory}{destroy}
\SetKwProg{Fn}{Function}{:}{}
\SetKw{KwTo}{to}
$remain\_num \gets coroutine\_pool.size$\;
\;
\Fn{\FScheduler{$coroutine\_pool$}}{
    \While{$remian\_num \neq 0$}{
        \For{\( i \gets 1 \) \KwTo  $coroutine\_pool.size$}{
            $coroutine\_pool[i].$\Fresume{}\;
            \If{$coroutine\_pool[i].$\Ffinish{}}{
                $coroutine\_pool[i].$\Fdestory{}\;
                $remain\_num \gets remain\_num - 1$
            }
        }
    }
}
\;
\Fn{\FCoro{...}}{
    \textbf{// code segment 1} \;
    \If{$remain\_num \neq 1$}{
        co\_await \textbf{suspend\_always}\;
    }
    \textbf{// code segment 2} \;
}
\end{algorithm}

% \blackcircleone We discuss graph computation tasks first.
\etitle{Graph Computation Tasks}.
For a specific graph computation task, it is necessary to analyze the patterns of vertex and edge access to determine the most suitable strategy.
As we have described in the example of interleaving execution mode in \autoref{fig:overlap_interleave}, each coroutine in a graph processing task represents a unit of concurrent execution that processes the subgraph within the overall task.

\etitle{Graph Update Tasks.}
% We only discuss batch update tasks, since single update tasks cannot be supported by the interleaved execution mode.
In \oursys, we support both individual updates and batch updates, depending on the specific requirements of the application scenario.
However, it is important to note that the coroutine optimization scheme used in \oursys can only optimize batch edge updates.
Specifically, for individual updates, we apply vertex-level locks on the \ourstorage to ensure accuracy.
For batch updates,
similar to systems that support batch updates (\eg Aspen~\cite{dhulipala2019low} and Terrace~\cite{pandey2021terrace}), we classify update tasks by source vertex before updating to avoid the overhead of locks caused by data conflicts.
Each vertex that is waiting to be updated will have a collection of update tasks.
The task allocator models each vertex's task collection as a coroutine, which is similar to \FE in \autoref{alg:get_e}.
Since insertions may modify the structure making it difficult to estimate the number of suspension points before the update, we opt for the trimmed-polling scheduler for batch updates.

% \textbf{Structural design considerations for better load balancing.}
% --------------------------for response--------------------------
% \textbf{Takeaways for Co-Design in }\ourstorage.
% We design the data structures for better coroutine load balancing.
% The design of the GTChain allows for a finer-grained sub-task division and also enables the assignment of tasks across vertices. 
% The choice of the B+ tree enables \PrOneNei because it can accurately know the number of suspension points required before computation, facilitating more precise load balancing for coroutines.
% --------------------------for response--------------------------

\subsection{Hybrid Prefetching}
% \label{sec:adapt-layer}

Hardware prefetching fetches data silently to users.
However, if hardware prefetching succeeds, it will lead to additional overhead to repeatedly operate software prefetching.
Accordingly, the ideal scenario is to use software prefetching when the hardware prefetching fails to fetch data and to avoid using software prefetching when hardware prefetching is effective.

Assume that the cost of a cache miss is $C_{m}$ and the probability of a cache miss being resolved by hardware prefetching is $P_{h}$.
This probability varies based on the memory layout of the data.
The software prefetching solution via coroutines incurs a certain overhead, $C_{coro}$.
If the cost $C_{m} \times (1 - P_{h}) \textless C_{coro}$, then %using 
the software prefetching becomes redundant. %becomes counterproductive.
To avoid this overhead of redundant data fetching, we aim to avoid employing software prefetching on data that is highly likely to be prefetched by the hardware prefetchers.
We also need to consider the phenomenon caused by the skewed graph structures from three perspectives discussed in \autoref{subsec:cache_miss_and_hard_pre}.
% -----------------------------for response--------------------------------
% lhf-mark: 为了篇幅 删除
% We consider the phenomenon caused by the skewed graph structures from three perspectives:
% \ding{172} Array-like structures offer good performance for computation through hardware prefetching, but the frequent update requirements can limit their use.
% \ding{173} Hardware prefetching can also be effective on linked list-like structures, but fetching the front few blocks of a linked list will cause cache misses in theory.
% When it comes to the graphs where most vertices are low-degree (\ie the linked lists are short), hardware prefetching has no advantage on most linked lists, thus dramatically increasing the time cost.
% \ding{174} Maintaining ordered data is also necessary for graph queries and GPM, so tree-like structures are needed.
% However, hardware prefetching does not work effectively on them.
% -----------------------------for response--------------------------------
%lhf:预取同一棵树的时候，何时需要放弃软件预取？（实现思路：通过时间戳？LRU？可能只为度数Top的Hub Vertex设计？）
% Thus, we choose to design hybrid strategies for specific patterns to perform better, as shown in \autoref{fig:hybrid_strategy}. 

\begin{figure}[t]
  \centering
  \includegraphics[width=\linewidth]{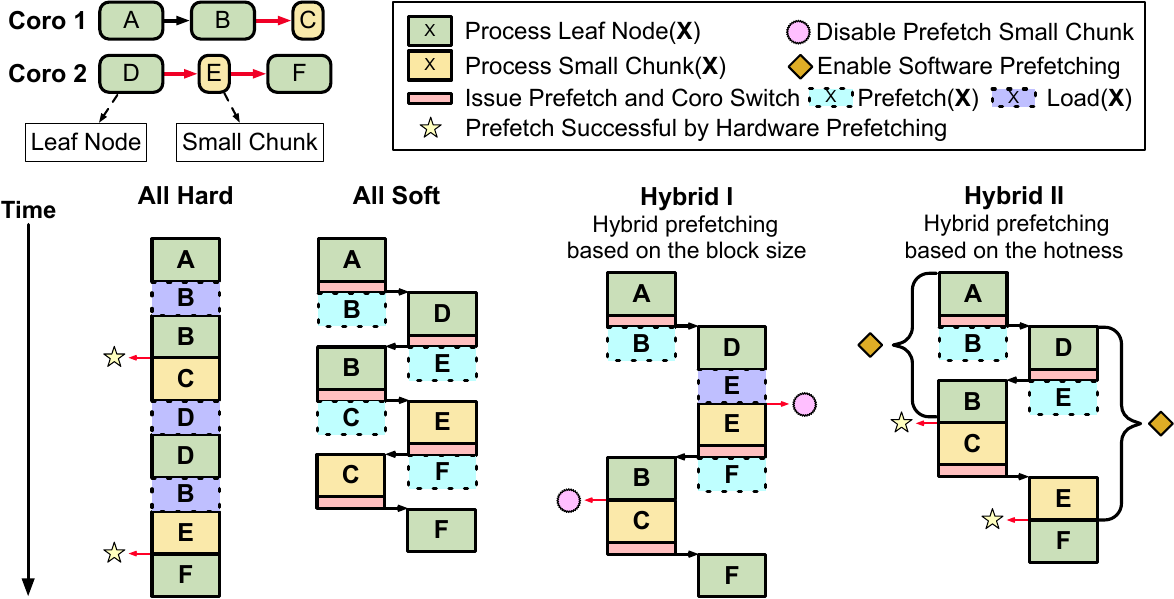}
  \caption{Hybrid prefetching.}
  \label{fig:hybrid_strategy}
  \vspace{-0.2in}
\end{figure}

Therefore, we aim to use software prefetching to complement hardware prefetching for better performance.
However, implementing software prefetching via coroutines also has overhead, as shown in \autoref{sec:exe-layer}.
To better utilize hardware and software prefetching, we design four prefetching strategies, as shown in \autoref{fig:hybrid_strategy}.

\emph{All Hard.}
This strategy uses hardware prefetching exclusively.
The performance of applying \emph{All Hard} depends on how much the data structure supports hardware prefetching.
In \ourstorage, designs like cache alignment, B+ trees, and GTChain aim to support hardware prefetching better.

\emph{All Soft.}
This strategy uses software prefetching exclusively.
Although it may conflict with hardware prefetching in various tasks, \autoref{subsec:ab-performance} proves it is an effective strategy.

\emph{Hybrid strategy I (Hybrid prefetching based on the block size).}
In the implementation of \ourstorage, all small chunks are arranged contiguously in memory, resulting in a high probability of being prefetched by the hardware prefetcher.
In this case, we skip software prefetching for small chunks and instead rely on the hardware prefetcher’s inherent capability to prefetch these small chunks.
In most cases, this strategy is more effective than the previous ``All Soft'' strategy.
This strategy is more suitable for scenarios where small chunks have a high cache hit rate.
% Examples include machines with larger caches or situations with many low-degree vertices where computation tasks involve sequential data access.

% This strategy is proposed based on the jump-pointer mechanism in theory ~\cite{DBLP:series/synthesis/2014Falsafi}.
% The strategy uses software prefetching for the first $N$ blocks when traversing the chain, and switches to hardware prefetching from the $(N+1)$-th block onwards.
% Its advantage is that it can achieve optimal performance under ideal conditions (as shown in \autoref{fig:hybrid_strategy}, where \emph{Hybrid I} perfectly predicts the failure timing of hardware prefetching).
% However, in practice, due to cache replacement policies, cache coherency protocols, and other factors, ensuring the effectiveness of hardware prefetching and selecting the appropriate parameter $N$ is challenging. 
% Therefore, the performance stability of the strategy may be relatively poor.
% \emph{Hybrid I} could be applied to graph computation tasks involving sequential data access.

\emph{Hybrid strategy II (Hybrid prefetching based on the hotness).}
To further prevent redundant data fetching, we consider the cold start issue of the jump-pointer mechanism described in Section 2.2.
Based on Hybrid I, we use software prefetching at the beginning of the linked list.
For the latter part of the linked list, we solely rely on the hardware prefetcher’s jump-pointer mechanism.
In summary, this strategy utilizes software or hardware prefetching based on the cache's ``cold'' and ``hot'' data status.
Under ideal conditions, this strategy achieves optimal results.
However, it requires iterative tuning and adjustments to determine the optimal threshold parameters of using hardware or software prefetching based on the cache's state.
Thus, the strategy is particularly suitable for scenarios where repeated fine-tuning of computations is possible.

\section{EVALUATION}
\label{expr}

This section introduces the experimental setup, and then validates the efficiency of the proposed system through various evaluations.

\vspace{-0.1in}

\subsection{Experimental Setup}

We run the experiments on a server equipped with one Intel Xeon Platinum 8269CY CPU whose clock speed is 2.5GHz. 
The server contains 26 cores (52 hyper-threads) and 371GB of main memory.
The sizes of the L1-cache, L2-cache, and L3-cache of the CPUs are 832KB(data cache)+832KB(instruction cache), 26MB, and 35.8MB, respectively.
We compiled all systems with the O3 optimization flag using GCC v10.3. All experiments are conducted with 52 worker threads by default unless otherwise stated, with each test running five times to report the average outcome.
% Each instance is executed $5$ times and the the average is reported.
% Experimental data is obtained by calculating the median from the results of five rounds of runs.

\noindent\textbf{Datasets.} 
We used five real-life graphs in our evaluation (see \autoref{tab:dataset}), including social networks Livejournal~\cite{Livejournal_dataset}, Hollywood~\cite{hollywoodUK} and Com-friendster~\cite{OrkutComFriend}, and web graph UK-2002~\cite{hollywoodUK} and Orkut~\cite{OrkutComFriend}.
The detailed information of the datasets is shown in \autoref{tab:dataset}, which includes the name of the dataset, the number of vertices, the number of edges, and average degrees. 
To simulate the realism of dynamic graph processing scenarios in practice, the datasets are shuffled. This is because other structures that contain pointers may be penalized during loading due to the inherent order of the data.
To ensure fairness, all systems load datasets using the weighted graph mode, and random weights are generated for those unweighted datasets.
Besides, we used the SNB-sf1000 dataset \cite{DBLPdel} to test the system's update performance in real-world scenarios. The SNB-sf1000 dataset includes timestamped person nodes and their relationships, along with 7,285 vertex deletions and 39,877,751 edge insertions. 
We also used the Ogbl-citation2 \cite{ogbl} dataset, which includes 128-dimensional word2vec features, to test graph attribute updates.
% }

\noindent\textbf{Workloads.} 
We conduct three types of workloads. (i) Graph Query: we randomly query 5\% edges of the data set. (ii) Graph Algorithms: we use five typical graph analysis algorithms, \ie BFS, Single Source Shortest Path (SSSP)~\cite{DBLP:journals/tpds/ChakaravarthyCM17}, PageRank (PR)~\cite{page1998winograd}, Connected Components (CC) and Label Propagation (LP). (iii) Graph Update: we synthetically generate 10,000 random graph updates in the form of edge insertions and edge deletions.

\noindent\textbf{Competitors.}
We compare {\oursys} with state-of-the-art solutions including LLAMA \cite{DBLP:conf/icde/MackoMMS15}, PCSR \cite{DBLP:conf/hpec/WheatmanX18}, LiveGraph \cite{zhu2019livegraph}, GraphOne \cite{DBLP:journals/tos/KumarH20}, RisGraph \cite{DBLP:journals/corr/abs-2004-00803}, Teseo \cite{de2021teseo}, Sortledton \cite{fuchs2022sortledton} and Terrace \cite{pandey2021terrace}.
% 其中，LLAMA和PCSR都采用了CSR类似的结构以提升数据局部性。LiveGraph采用了日志结构以追加的方法插入数据以获取更快的插入性能。GraphOne和RisGraph分别采用了块状邻接链表和邻接数组来支持数据的快速插入。Teseo采用树状结构来存储边数据以支持快速的插入并保持邻居的有序存储. Sortledton和Terrace根据顶点数据来分别采用不同的结构存储邻居数据，以获取插入性能和图分析性能之间的平衡。

LLAMA, Teseo, and PCSR utilize structures similar to CSR to enhance data locality. LiveGraph employs a log-structured approach for data insertion to achieve faster insertion performance.
GraphOne and RisGraph adopt block-based adjacency lists and adjacency arrays, respectively, to support rapid data insertion. 
% Teseo~\cite{de2021teseo} uses a tree-based structure for storing edge data to facilitate quick insertions while maintaining the ordered storage of neighbors. 
Sortledton and Terrace employ different structures for storing neighbor data based on vertex degree, striking a balance between insertion performance and graph analysis capabilities.

\begin{table}[tbp]
    \caption{Description of graph datasets.}
    \vspace{-0.1in}
    \label{tab:dataset}
    \begin{tabular}{llll}
        \toprule
        Graph & \#V & \#E & $\overline{D}$ \\
        \midrule
        Livejournal~\cite{Livejournal_dataset} &         $4$,$846$,$609$ & $68$,$475$,$391$ & $14.12$ \\
        UK-2002~\cite{hollywoodUK} &           $18$,$484$,$117$ & $298$,$113$,$762$ & $16.13$ \\
        Com-friendster~\cite{OrkutComFriend} &    $65$,$608$,$366$ & $1$,$806$,$067$,$135$ & $27.53$ \\
        Orkut~\cite{OrkutComFriend} & $3$,$072$,$441$ & $117$,$185$,$083$ & $38.14$ \\
        % UK-2005 &           $39,454,746$ & $936,364,282$ & $23.73$ \\
        Hollywood~\cite{hollywoodUK} &         $1$,$139$,$905$ & $116$,$050$,$145$ & $101.81$ \\
        SNB-sf1000~\cite{DBLPdel} &         $3$,$144$,$492$ & $202$,$282$,$791$ & $64.33$ \\
        Ogbl-citation2~\cite{ogbl} &         $2$,$927$,$963$ & $30$,$561$,$187$ & $10.44$ \\
    \bottomrule
    \end{tabular}
    \vspace{-0.1in}
\end{table}

\subsection{Performance of Graph Query}
\label{subsec:graph-query}
% 我们评估了\oursys的执行边查询时间在四个数据集上，其在社交网络分析、推荐系统、交通网络等多个领域扮演着关键角色，例如在社交网络中，它可以用来检查两个用户之间是否存在直接的朋友关系。
% \ys{
We first evaluate the performance of edge queries by \oursys on all datasets in Table~\ref{tab:dataset}, which plays a crucial role in multiple domains such as social network analysis, recommendation systems, and transportation networks. 
For example, in social networks, edge queries can be utilized to check whether there is a direct friendship between two users or not.
% }

% \ys{
% 对于边的查询，\oursys在所有数据集上均胜于其它所有系统。
Figure \ref{fig:query} shows the normalized time for each system to execute queries on different datasets.
Here the response time of \oursys is treated as the baseline, \ie finishes in unit time. 
We can observe that \oursys outperforms others in all the cases.
Specifically, {\oursys} achieves an average 
13.7$\times$ (up to 16.8$\times$) speedup over Terrace, 
7.4$\times$ (up to 12.0$\times$) speedup over LLAMA, 
2.8$\times$ (up to 7.2$\times$) speedup over PCSR,
10.9$\times$ (up to 20.4$\times$) speedup over GraphOne,
4.7$\times$ (up to 8.5$\times$) speedup over RisGraph,
6.0$\times$ (up to 9.1$\times$) speedup over LiveGraph,
4.6$\times$ (up to 6.2$\times$) speedup over Teseo,
and 5.6$\times$ (up to 7.6$\times$) speedup over Sortledton.
% 其中对于Terrace、LLAMA和GraphOne在Hollywood数集上超过了10倍。这是由于Hollywood数据集的顶点的平均度最大(即101.81),对于高度顶点这三个系统的存储结构在查询时容易导致较大的查询开销。具体而言，Terrace采用了分层的结构，而当顶点的度数较大时，其邻居将会分布在多个层的数据结构中，这导致查询时需要在多个存储结构中进行查找。对于LLAMA通过构建多版本的数组来存储图更新，然而随着图更新的增加，其将会生成大量版本的数组，对于度数较大的顶点，其边可能存储在多个数组中，因此导致了查询性能低。对于GraphOne其采用了块中邻居链表来存储每个顶点的邻居，因此当顶点度数较大时，其需要遍历的链也越长，而导致较长的查询时间。
% 相比于其它系统而言，\oursys也取得了较高的加速比，这主要是因为\oursys采用了？？？
For Terrace, LLAMA, and GraphOne, they experienced a slowdown exceeding 10$\times$ on the Hollywood dataset. 
This can be attributed to the dataset's vertices having the highest average degree (\ie 101.81), which leads to significant query overhead for vertices with high degrees due to the storage structures of these three systems during retrieval.
Specifically, Terrace adopts a hierarchical structure. When the degree of a vertex is large, its neighbors will be distributed in multiple layers of data structures, which results in the need to search across multiple storage structures when querying.
% For LLAMA, it stores graph updates by constructing multi-version arrays. However, as the graph updates increase, a significant number of versions of the array will be generated. For vertices with a large degree, their edges may be stored in multiple arrays, thus resulting in lower query performance.
For GraphOne, it uses a block-based adjacency list to store the neighbors of each vertex. Therefore, for vertices with a high degree, this necessitates traversing a longer chain, thereby extending the query time.
Compared with other systems, \oursys also achieves better performance, mainly because \oursys adopts more ``stubby'' structures, \ie B+ Tree, as part of our edge storage.
The B+ tree, due to its strict rebalance rules, has fewer levels, meaning that we encounter fewer pointers and experience fewer cache misses during queries.
We can offset the benefits with software prefetching via coroutines to amortize the rebalance overhead during updates.
% }

\begin{figure}
% \vspace{-0.1in}
  \centering
  \includegraphics[width=\linewidth]{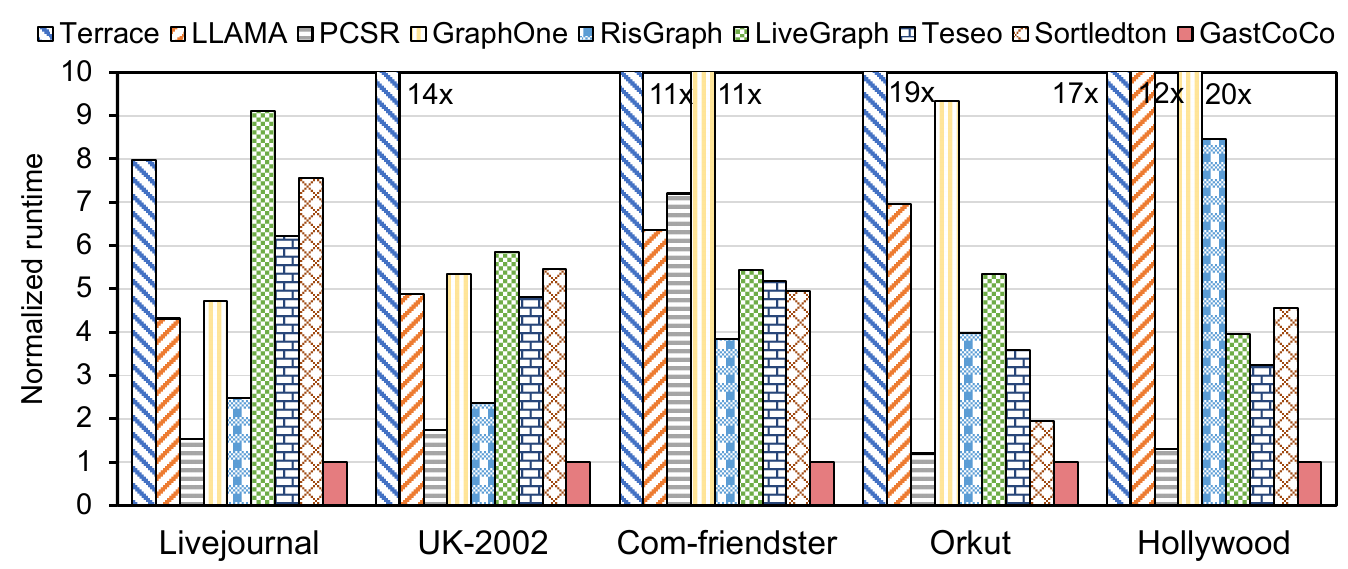}
  \vspace{-0.15in}
  \caption{Query execution time.}
  \label{fig:query}
  % \vspace{-0.1in}
\end{figure}

\begin{figure}[tbp]
% \vspace{-0.15in}
  \centering
    % \subfloat[Livejournal]{\label{fig:alg-LJ}
    %     \includegraphics[width=0.48\linewidth]{fig-sec6/alglj.pdf}
    % }
    % \subfloat[Com-friendster]{\label{fig:alg-CF}
    %     \includegraphics[width=0.48\linewidth]{fig-sec6/algcf.pdf}
    % }
    % \\
    % \vspace{-0.1in}
    % \subfloat[Orkut]{\label{fig:alg-ok}
    %     \includegraphics[width=0.48\linewidth]{fig-sec6/algok.pdf}
    % }
    % \subfloat[Hollywood]{\label{fig:alg-HW}
    %     \includegraphics[width=0.48\linewidth]{fig-sec6/alghw.pdf}
    % }
    \includegraphics[width=\linewidth]{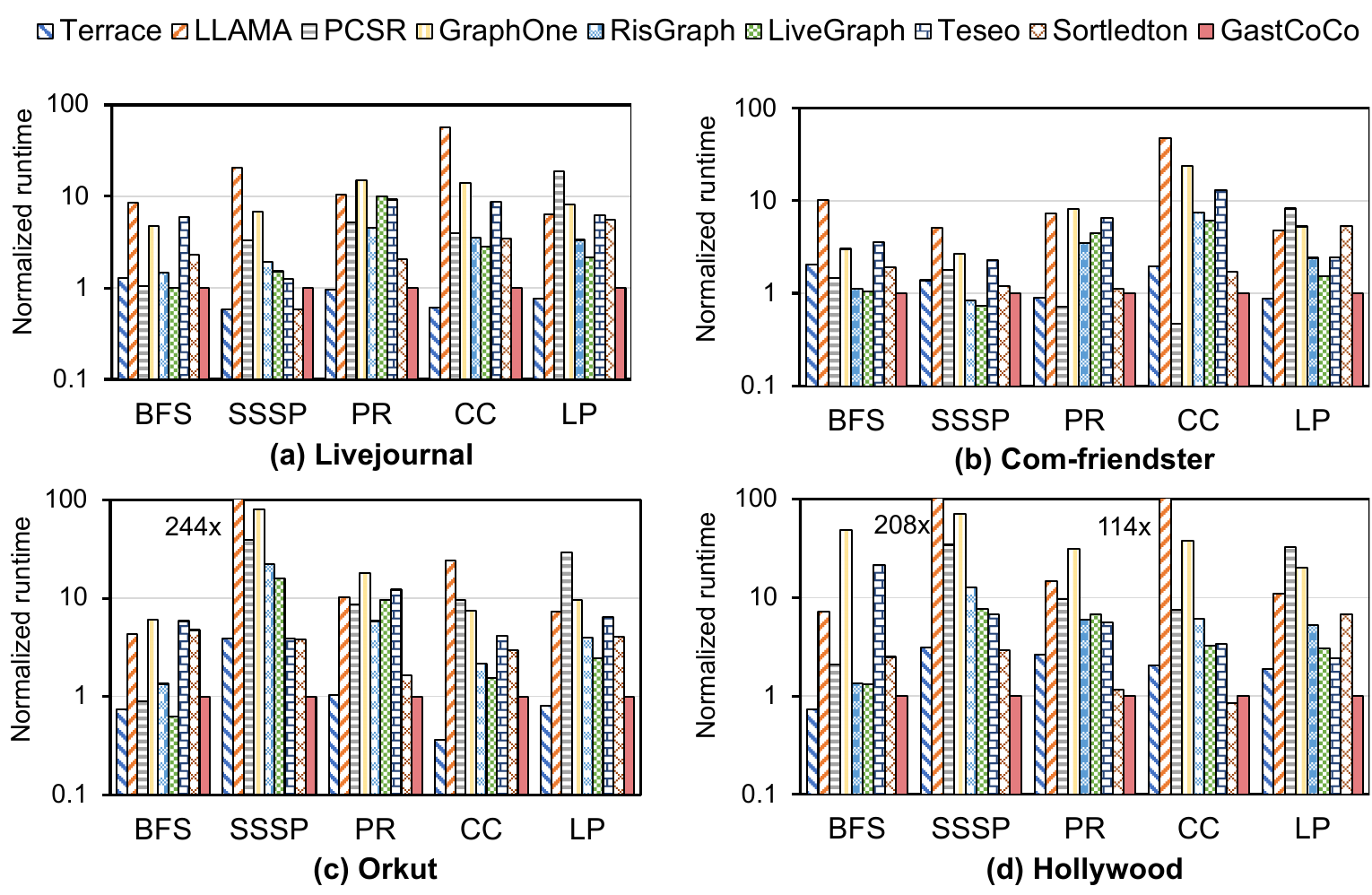}
    \vspace{-0.15in}
  \caption{The execution time of graph algorithms on different datasets.}
  \label{fig:runtime}
  % \vspace{-0.2in}
\end{figure}

\subsection{Performance of Graph Analysis}
\label{subsec:graph-alg}
% \ys{
We next evaluate the performance of graph analysis, including the execution time of five graph analytics algorithms, by comparing \oursys with competitors.
Figure \ref{fig:runtime} shows the normalized time for each system to execute graph analysis algorithms on different datasets.
Here the response time of \oursys is treated as the baseline, \ie finishes in unit time. 
We can see that \oursys outperforms others in most cases.
Specifically, {\oursys} achieves an average 
1.4$\times$ (up to 3.9$\times$) speedup over Terrace, 
41.1$\times$ (up to 243.8$\times$) speedup over LLAMA, 
11.0$\times$ (up to 39.7$\times$) speedup over PCSR,
21.7$\times$ (up to 79.4$\times$) speedup over GraphOne,
4.9$\times$ (up to 22.3$\times$) speedup over RisGraph,
4.2$\times$ (up to 16.0$\times$) speedup over LiveGraph,
5.8$\times$ (up to 21.7$\times$) speedup over Teseo, and
2.8$\times$ (up to 6.8$\times$) speedup over Sortledton.
% 从图\ref{fig:runtime}中可以看到LLAMA和GraphOne表现出了最大的slowdown。
As shown in Figure \ref{fig:runtime}, LLAMA and GraphOne exhibit the most significant slowdown.
% 对于LLAMA其采用了多版本数组来存储顶点的邻居边，随着数据的插入，一个顶点的邻边可能会分布在多个不同的版本数组中，这对于图分析算法非常不利，因为图分析算法总是需要访问一个顶点的全部邻居。LLAMA通过指针一个顶点分布在不同版本数组的边连接起来，随着减少了查找的时间，然而这依然导致了大量的指针追逐。对于GraphOne它同样遭遇了昂贵的指针追逐的影响，由于其采用了块状邻接链表。
LLAMA employs a multi-version array to store the neighboring edges of vertices. As data is inserted, the edges of a vertex may be distributed across multiple arrays, which is disadvantageous for graph analysis algorithms since they always require access to all neighbors of a vertex. 
LLAMA connects the edges of a vertex distributed in different arrays by pointers, thereby reducing search time. However, this still results in extensive pointer chasing. GraphOne faces similar issues with expensive pointer chasing due to its use of block-based adjacency lists.
% 此外从图\ref{fig:runtime}可以观察到Terrace、PCSR和Sortledton获得了较好的图分析性能。这是由于Terrace和Sortledton采用了数组来存储小度顶点的邻居，PCSR采用数组存储全部顶点的邻居，这导致对于空间局部性较好。然而，这种设计也导致了他们的更新成本更高由于数据插入而导致的数据挪动，这将在后面的实验\ref{expr:insert}中详细介绍。此外，Terrace和Sortledton采用了跳表或者B-Tree来存储大读顶点的邻居，这依然会导致大量的指针追逐。相比较而言，\oursys采用了软件预取和协程来加速图分析中的数据读取并减少cache miss，从而获得了更好的图分析性能。
Furthermore, Figure \ref{fig:runtime} shows that Terrace, PCSR, and Sortledton achieved better graph analysis performance compared to other competitors. 
This is attributed to Terrace and Sortledton using arrays to store the neighbors of vertices with low degrees, and PCSR using arrays to store the neighbors of all vertices, which have a better spatial locality. 
However, this design also results in higher update costs due to data movements caused by insertions/deletions, which will be detailed in Section \ref{expr:insert}.
Additionally, Terrace and Sortledton employ B-Trees or skip lists to store the neighbors of high-degree vertices, still leading to extensive pointer chasing. 
In contrast, \oursys utilizes software prefetching and coroutines to accelerate data retrieval in graph analysis and reduce cache misses, thereby achieving superior graph analysis performance.

\begin{figure}
% \vspace{-0.2in}
  \centering
  \includegraphics[width=0.8\linewidth]{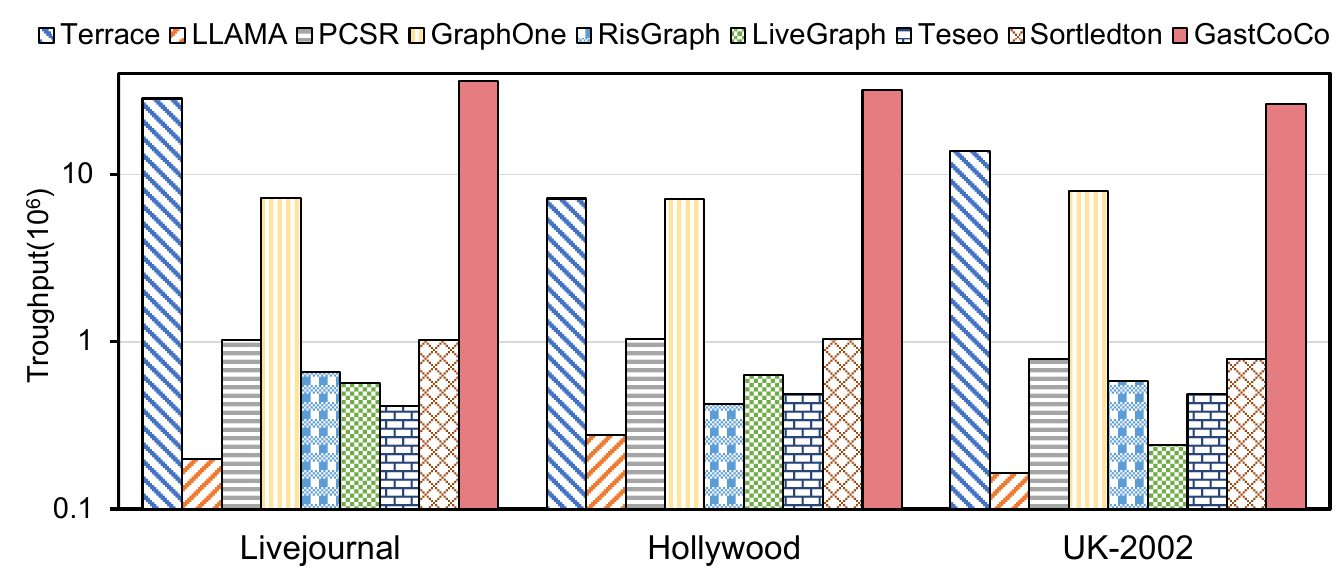}
  % \vspace{-0.01in}
  \caption{
  %10000 batch insert on different dataset.
   The throughput of graph updates.
  }
  \label{fig:edge_insert}
  % \vspace{-0.1in}
\end{figure}

\begin{figure}
% \vspace{-0.1in}
  \centering
    \includegraphics[width=\linewidth]{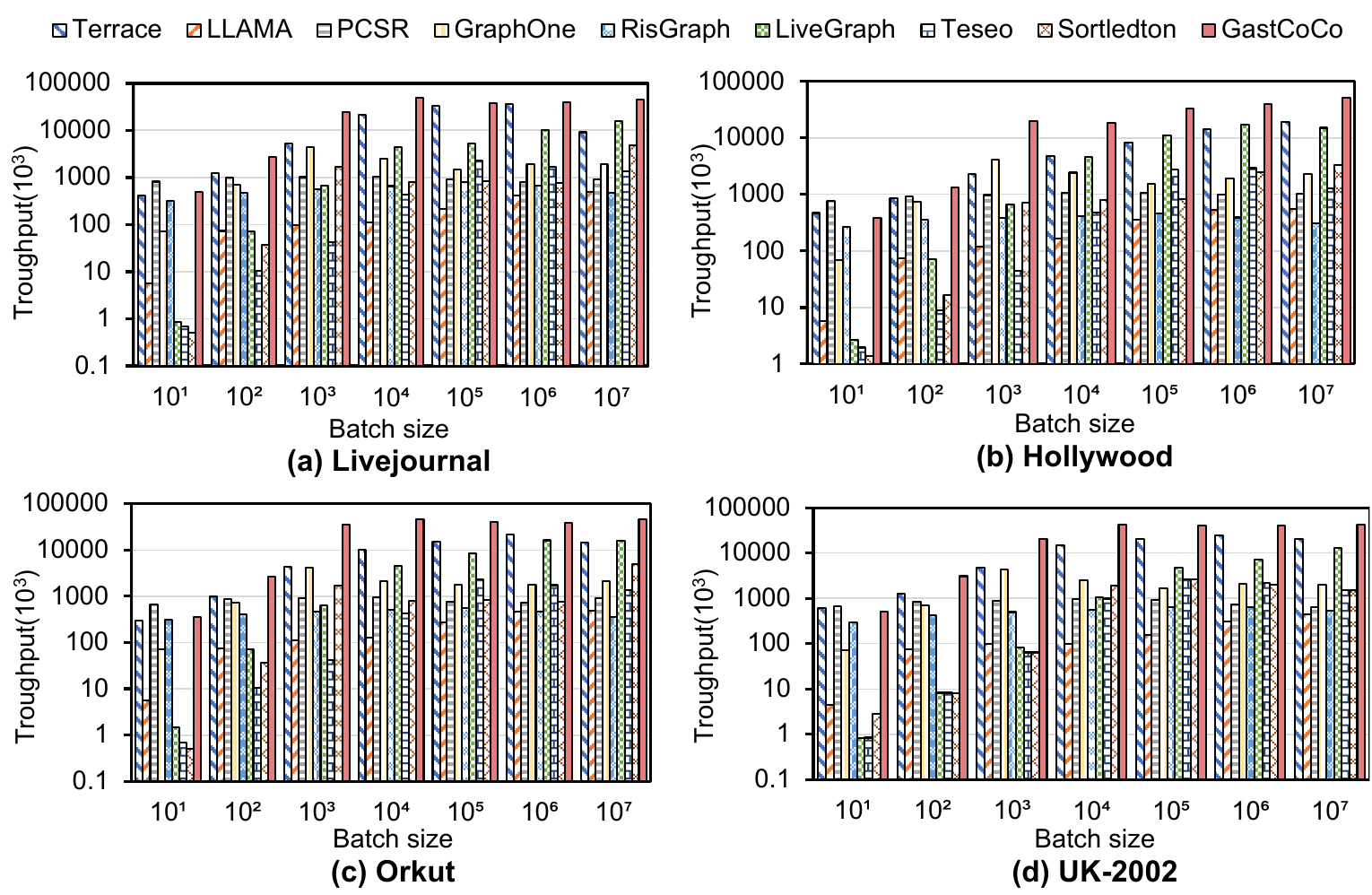}
    \vspace{-0.15in}
  \caption{
  % Hollywood Batch Insert 16 thread.
   The throughput of graph updates with varying batch sizes on different datasets.}
  \label{fig:batch_size}
  % \vspace{-0.1in}
\end{figure}

% \begin{figure*}[ht]
%   \centering
%     \subfloat[16 thread]{\label{}
%     \includegraphics[width=0.5\linewidth]{fig-sec6/edge16UK.pdf}}
%     \subfloat[52 thread]{\label{}
%     \includegraphics[width=0.5\linewidth]{fig-sec6/edge52UK.pdf}}
%   \caption{UK2002 Batch Insert.}
%   \label{fig:alg-and-cm}
% \end{figure*}

\subsection{Performance of Graph Updates}
\label{expr:insert}

% \ys{
% 我们评估了不同系统在LiveJournal,Hollywood和UK-2002上插入性能。如图 \ref{fig:edge_insert} 显示所有系统在不同数集下每秒百万边缘 (meps) 的吞吐量，很显然\oursys 在所有测试中均优于所有竞争对手。
We evaluate the graph update performance of different systems on LiveJournal, Hollywood and UK-2002 for continuous edge insertions/deletions of 10K edges.
Figure \ref{fig:edge_insert} shows the throughput (edges per second) for all systems under different data sets. \oursys outperforms all competitors in all tests.
Specifically, {\oursys} achieves an average 
2.5$\times$ (up to 4.5$\times$) speedup over Terrace, 
152.4$\times$ (up to 180.2$\times$) speedup over LLAMA, 
33.3$\times$ (up to 35.4$\times$) speedup over PCSR,
4.3$\times$ (up to 5.0$\times$) speedup over GraphOne,
58.2$\times$ (up to 75.0$\times$) speedup over RisGraph,
74.4$\times$ (up to 109.4$\times$) speedup over LiveGraph,
68.8$\times$ (up to 87.2$\times$) speedup over Teseo,
and 33.3$\times$ (up to 35.4$\times$) speedup over Sortledton.
% 从中可以看到Terrace、GraphOne和\oursys具有更高的吞吐量。对于GraphOne，它采用了块状链表的存储结构，这是非常利于插入的相对于连续存储结构。然而，正如前面前面实验所展示的，它在图查询和图分析上表现较差。
% 对于Terrace和\oursys都采用了批量插入的设计，即在更新之前按源顶点对更新任务进行分类，为了避免数据冲突引起的锁开销。此外，由于Terrace保持插入的数据有序的存储，因此插入前需要进行查找以确定插入数据的位置，查找的过程中将会导致大量的cache miss。相对于而言，\oursys虽然也需要保持有序的插入，但是我们采用了协程相关的设计有效的减少了cache miss, 因此获得了比Terrace更好的性能。
It can be observed that Terrace, GraphOne, and \oursys have higher throughput. GraphOne utilizes a block-based adjacency list storage structure, which is particularly advantageous for insertions. 
However, as demonstrated in the previous experiment, it performs poorly in graph queries and graph analysis.
Both Terrace and \oursys adopt a batch updates design, \ie classify update tasks by source vertices before updating, to avoid lock overhead caused by data conflicts.
Moreover, since Terrace maintains the updated data in an ordered manner, a search operation is required to determine the position of the new data, leading to a significant number of cache misses. 
In contrast, although \oursys also maintains ordered updates, our coroutine-related design effectively reduces cache misses, thereby achieving better performance than Terrace.
% }

\eat{
\begin{figure}
%\vspace{-0.1in}
  \centering
  \includegraphics[width=\linewidth]{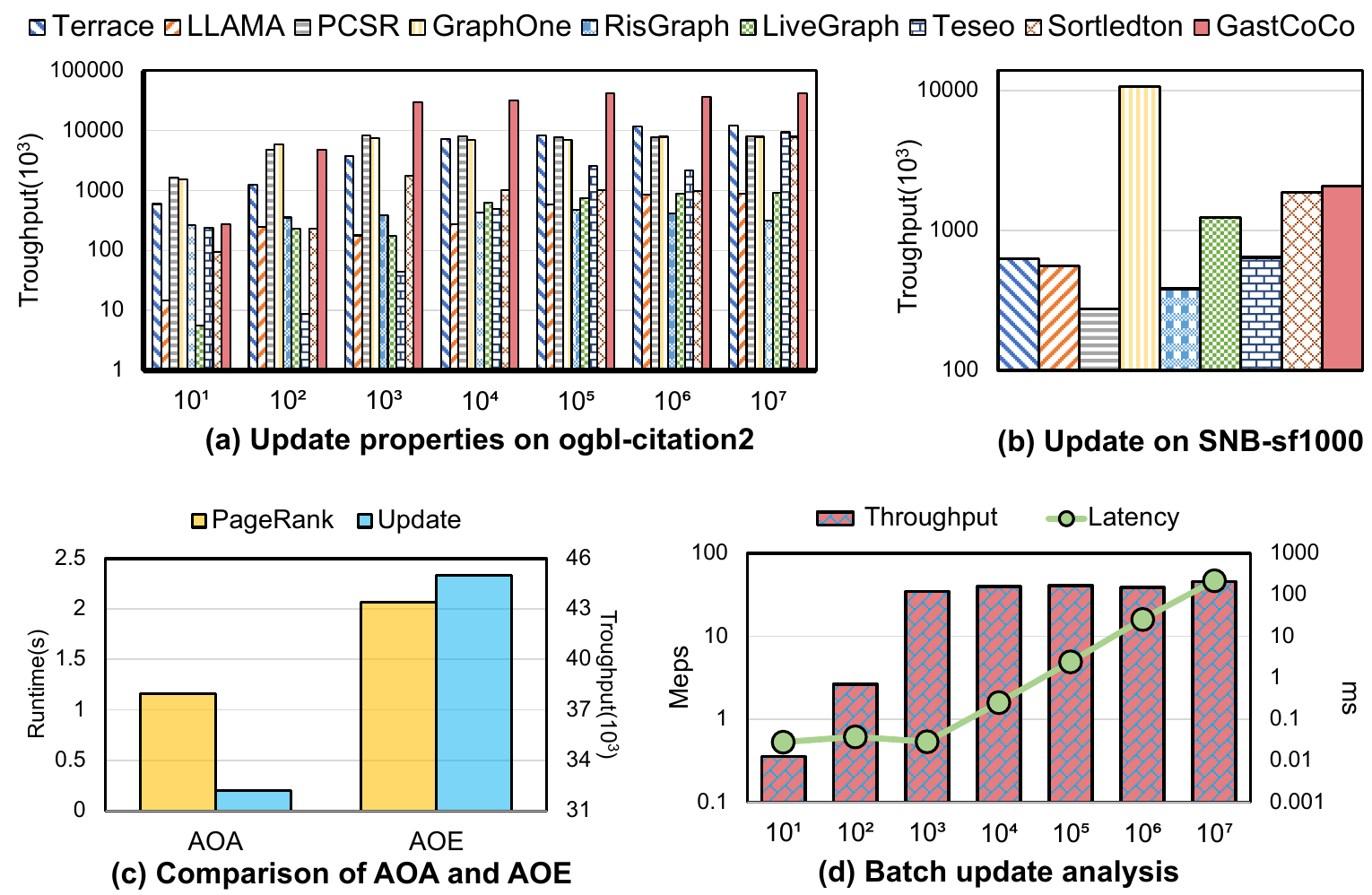}
  \vspace{-0.3in}
  \caption{\ys{XXXXXXXXXXXXXXXXXX}}
  \label{fig:query}
  \vspace{-0.15in}
\end{figure}
}

% 正如您所指出的，由于 OGB 数据集的属性较大，因此在一定程度上影响了 \oursys 的性能。此外，由于更新属性不需要结构更改，因此许多系统在此任务中表现出色。对于大属性图数据，使用 AOE 更有利于更新，因为每条边的属性与目标顶点 ID 具有更好的空间局部性。相反，对于不需要属性的图算法（例如 PageRank），使用 AOA 更有利，因为相邻边的目标顶点 ID 之间的空间局部性更好。我们在 \red{图 15(c)} 中显示的实验证实了这一结论。此外，不同的系统以各种方式实现边属性。为了确保公平比较，所有系统都使用加权图模式加载数据集（即使对于非加权图数据集，我们也会随机生成边权重）。我们已将此描述附加到 \textbf{\red{第 7.1 节，数据集}}。因此，在我们之前的更新实验中，属性存储问题已被考虑在内。

% 我们评估了不同系统更新属性的性能。具体而言，我们在OGB \cite{}数据集上测试了不同系统更新
% \ys{Property updates}.

% In each column, we provide the system's throughput and its ranking in that column. 
% From \autoref{tab:batch_insert_LJ} and \autoref{tab:batch_insert_HW}, we can see that our system ranks first in insertion performance under most batch sizes. 
% In cases of smaller batch sizes, the performance is not as ideal due to the significant proportion of time taken for coroutine construction and scheduling in the overall insertion time. 
% When the batch size is 1E7, the performance declines due to the very large batch size and the massive amount of insertions for certain vertices. 
% This requires multiple insertions into the same vertex's adjacency structure. 
% Frequent insertions increase the likelihood of various structure components being in the cache, which can decrease the effectiveness of prefetching.

\subsection{Varying Batch Size of Graph Updates}
% \ys{
We also conduct experiments to evaluate the impact of batch size of graph updates on throughput performance.
Figure \ref{fig:batch_size} reports the throughput of the systems on graph updates with different batch sizes on LiveJournal, Hollywood, Orkut, and UK-2002.
For each batch size from $10$ to $10^7$, we report the throughput (edges per second) for all systems.
% 图\ref{fig:batch_size}显示了\oursys在batch size 从$10^3$到$10^7$的范围内击败了所有系统，主要的原因与第Section \ref{expr:insert}中的分析一样。此外当batch size大于$10^3$之后，\oursys的吞吐能够保持相对稳定，这表明了\oursys具有良好的扩展性。
% 在batch size为$10$和$100$时，Terrace、PCSR和GraphOne都超过了\oursys。
Figure \ref{fig:batch_size} shows that \oursys outperforms other systems with batch sizes ranging from $10^3$ to $10^7$, the reason for which is primarily consistent with the analysis provided in Section \ref{expr:insert}. Moreover, when the batch size exceeds $10^3$, the throughput of \oursys remains relatively stable, indicating that \oursys possesses excellent scalability.
On the other hand, when the batch size is $10$ and $100$, Terrace, PCSR, and GraphOne exceed \oursys.
% }
This is because the construction and scheduling overheads in \oursys's \ourstorage are difficult to balance against the benefits brought by coroutines for smaller batch sizes.

\begin{figure}[t]
% \vspace{-0.2in}
  \centering
    % \subfloat[Runtime of SSSP]{\label{fig:sssp-ab}
    %     \includegraphics[width=0.48\linewidth]{fig-sec6/sssp-ab.pdf}
    % }
    % \subfloat[Throughput of graph updates \todo{lhf: wrong}]{\label{fig:bi-ab}
    %     \includegraphics[width=0.48\linewidth]{fig-sec6/bi-ab.pdf}
    % }
    \includegraphics[width=\linewidth]{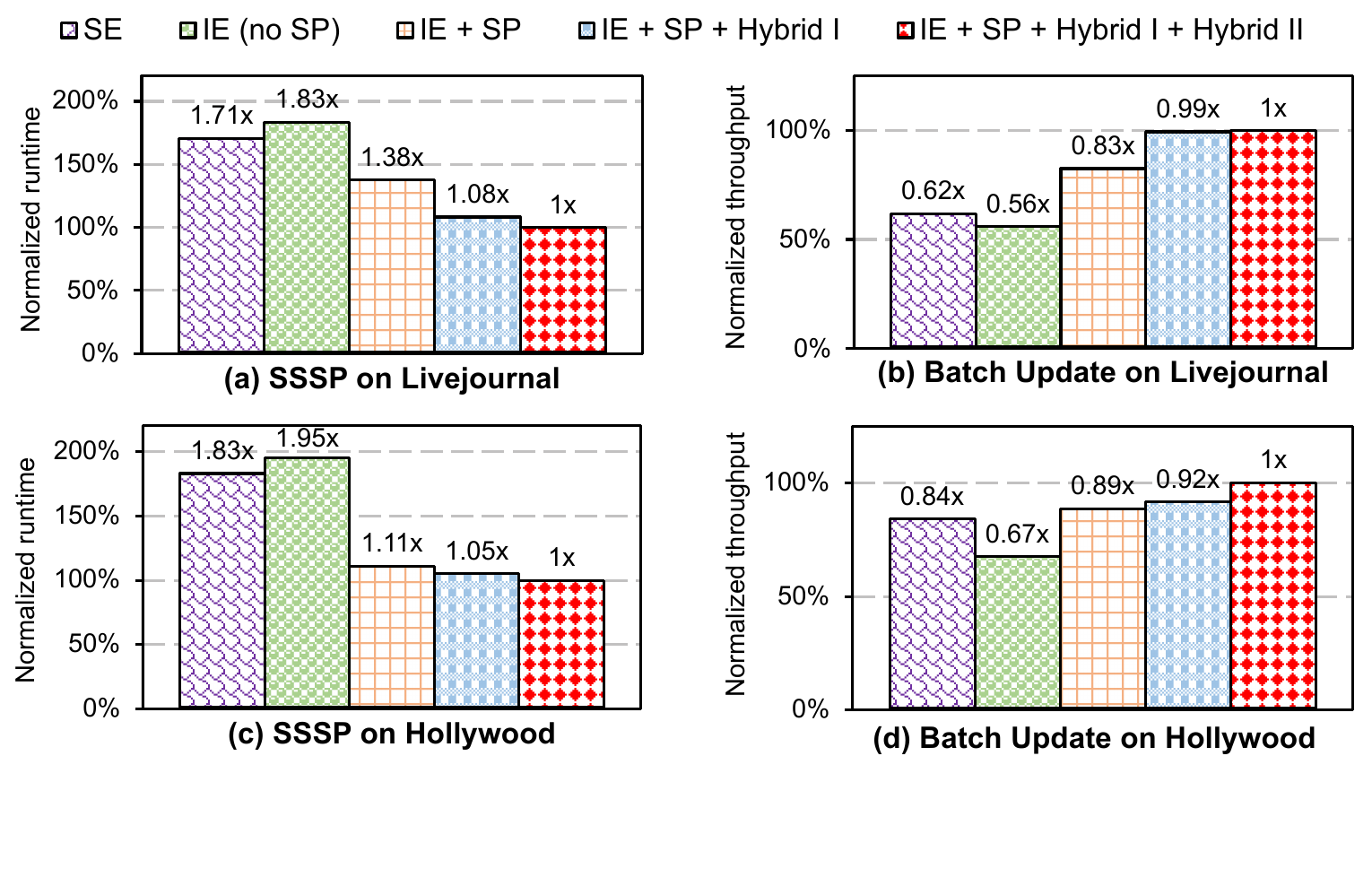}
    \vspace{-0.45in}
    
  \caption{Performance analysis. (SE represents Sequential Execution, IE represents Interleaved Execution, and SP represents Software Prefetching)}
  \label{fig:performance}
  % \vspace{-0.2in}
\end{figure}

% 我们进一步评估了所有系统的线程扩展性。图\ref{fig:thread}报告了所有系统在Livejournal上插入$10^4$条边加速比，即每个系统相对于自己使用一个线程时插入性能的加速比。
% 大部分系统在低于8个线程时，随着线程数的增加，加速比逐步增大。然而，当线程数超过8线程时或者16线程时加速比逐步下降。这是因为

% LLAMA 对于边插入的可扩展性很差，因为它将所有新边存储在写入存储中，而写入存储会受到线程的竞争。 

\begin{figure}
% \vspace{-0.05in}
\begin{minipage}{0.48\linewidth}
\centering
 \includegraphics[width=\textwidth]{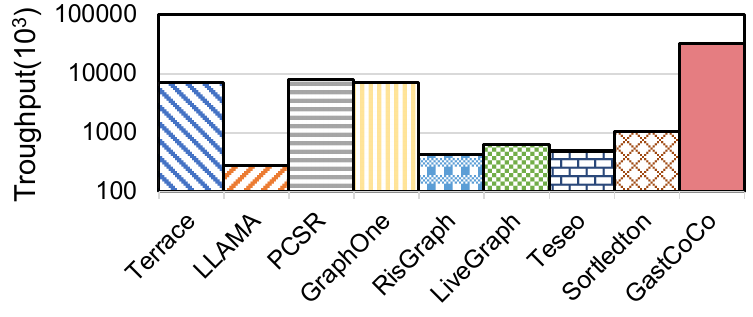}
 % \vspace{0.05in}
  \caption{Update properties on ogbl-citation2.}
  \label{fig:ogb}
\end{minipage}\hfill
\begin{minipage}{0.48\linewidth}
\centering
 \includegraphics[width=\textwidth]{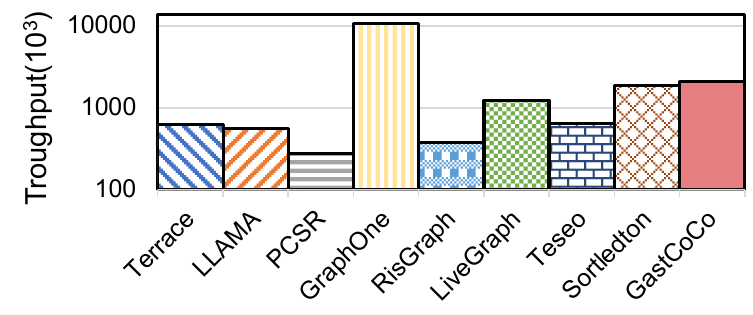}
  \caption{Update on SNB-sf1000.}
  \label{fig:snb}
\end{minipage}
\vspace{0.05in}
\end{figure}

% lhf-mark: 图17 18 废除，为了篇幅。
% \begin{figure}
% \begin{minipage}{0.48\linewidth}
% \centering
%  \includegraphics[width=\textwidth]{fig-sec6/aoaaoe.pdf}
%   \vspace{-0.3in}
%   \caption{Comparison of AOA and AOE on ogbl-citation2.}
%   \label{fig:aoa}
% \end{minipage}\hfill
% \begin{minipage}{0.48\linewidth}
% \vspace{-0.13in}
% \centering
%  \includegraphics[width=\textwidth]{fig-sec6/fresh.pdf}
%   \vspace{-0.22in}
%   \caption{Batch update analysis.}
%   \label{fig:fresh}
% \end{minipage}
% \vspace{-0.3in}
% \end{figure}

\subsection{Real-World Graph Updates}

Firstly, we conduct a properties update experiment using the OGB dataset.
The results are shown in \autoref{fig:ogb}.
Since updating properties does not cause structural changes, many systems have shown good performance in this task.
% lhf-mark: 为了篇幅删了下面3行
% For graph data with large properties, using AOE in \ourstorage is more beneficial for updates because each edge's property has better spatial locality with the destination vertex ID.
% Conversely, using AOA is more beneficial for graph algorithms that do not require properties (\eg PageRank), as the spatial locality between the destination vertex IDs of neighboring edges is better.
%% The conclusion could also be confirmed by our experiment shown in \red{\hyperlink{aoa}{Figure 16}}.
%% 即，在图算法((i.e., PageRank)上，AOA运行时间低于AOE，而在更新性能上，AOE获得了更高的吞吐量。
% The conclusion can also be confirmed by our experiment shown in \autoref{fig:aoa}, where AOA has a lower runtime than AOE in graph algorithms (i.e., PageRank), while AOE achieves higher throughput in update performance.

Secondly, we conduct a real-world graph update experiment on the SNB-sf1000 dataset to test the system's update performance in real-world scenarios. 
Unlike the previous update experiments, this dataset includes vertex deletions.
The SNB-sf1000 dataset includes timestamped person nodes and their relationships, along with 7,285 vertex deletions and 39,877,751 edge deletions. When deleting a vertex, we need to delete all its edges at the same time.
% 我们测试了在SNB-sf1000上不同系统执行的图更新的吞吐量，结果如图16所示，可以看到GastCoCo依然在优于大多数系统，除了GraphOne. For GraphOne, it utilizes a block-based adjacency list storage structure, which is particularly advantageous for insertions. However, as demonstrated by previous experiments, it performs poorly in graph queries and analysis.
We test the throughput of graph updates on different systems using the SNB-sf1000 dataset.
As shown in \autoref{fig:snb}, GastCoCo outperforms most systems except for GraphOne.
GraphOne uses a block-based adjacency list storage structure and does not actually delete content during updates, but rather appends deletion records, which is particularly advantageous for insertions.
However, as demonstrated by the experiments in \autoref{subsec:graph-query} and \autoref{subsec:graph-alg}, it performs poorly in graph queries and analysis. In contrast, our system shows good performance in both graph updates and analysis.

% lhf-mark: 删去7.7 为了篇幅
% \subsection{Trade-off between Throughput and Latency}

% \oursys supports both batch and individual edge updates.
% The granularity of batch updates can be adjusted according to specific applications.
% %The decision to use batch updates depends primarily on the updates nature provided by the upper-layer application.
% %If \oursys receives a batch of updates at any given time point, we opt for the batch update mode, which is accelerated by software prefetching via coroutine.
% Larger batch sizes can lead to higher throughput, which means that each edge must wait for more edges to be updated, resulting in data updates with lower freshness.
% To explain this, we conduct an experiment to measure the trade-off between throughput and data freshness, as shown in \autoref{fig:fresh}.

% As the batch size increases, throughput rises, but latency also increases.
% Additionally, at larger batch sizes, the increase in throughput becomes marginal.
% From \autoref{fig:fresh}, it is evident that when the batch size is $10^4$ or $10^5$, \oursys achieves high throughput while maintaining low latency.
% Therefore, unless the upper-layer application specifies a fixed batch size, \oursys will choose a batch size of $10^4$ or $10^5$ when actively initiating batch update mode.

\vspace{-4pt}
\subsection{Performance Analysis}
\label{subsec:ab-performance}

We conduct performance analysis on SSSP and batch insertions for our designs to understand the performance gains and losses.
SSSP includes \GVc, making the probability of hardware prefetching failures increase and software prefetching more effective.
Batch insertions utilize update/query pointers in \ourstorage for locating update positions in the edge storage structure.
Hence, the hybrid strategy I cannot be applied as it is only implementable on GTChain.
In particular, \textit{Sequential Execution} (SE) means executing tasks in sequential execution mode (refer to \autoref{subsec:interleave-exe});
\textit{Interleaved Execution} (IE) means using coroutines to form an interleaved execution mode (refer to \autoref{subsec:interleave-exe});
\textit{Software Prefetching} (SP) means incorporating software prefetching instructions during the interleaved execution process;
\textit{Hybrid I} and \textit{Hybrid II} represent employing hybrid strategy I with hardware failure time and hybrid strategy II with the content to be prefetched respectively in \autoref{sec:adapt-layer}.
From \autoref{fig:performance},
we could observe that the interleaved execution mode (IE no SP) spends longer running time compared with the sequential execution mode (SE) due to the additional switching cost of coroutines.
However, if we further employ software prefetching and hybrid prefetching strategy, \oursys achieves at most $1.83\times$ (graph computation) and $1.62\times$ (graph update) speedup compared with sequential execution on \ourstorage.

\begin{table}[tbp]
% \vspace{-0.1in}
\centering
\footnotesize
\setlength{\tabcolsep}{3pt}

\caption{Cache miss and cache stall percentage for different workloads and execution modes.}
\label{table:cache_miss}
\vspace{-0.1in}
\begin{tabular}{|l|c|c|c|c|c|}
\hline
\multirow{2}{*}{Workload} & \thead{Execution \\ mode} & \multicolumn{2}{c|}{Livejournal} & \multicolumn{2}{c|}{Hollywood} \\ \cline{3-6} 
  & & Cache miss & Percentage & Cache miss & Percentage \\ \hline
\multirow{2}{*}{PageRank} & SE & $2.09 \times 10^9$ & 50.4\% & $2.70 \times 10^9$ & 69.5\% \\ \cline{2-6} 
 & IE + SP & $2.01 \times 10^9$ & 42.9\% & $2.26 \times 10^9$ & 49.4\% \\ \hline
\multirow{2}{*}{SSSP} & SE & $1.75 \times 10^9$ & 31.7\% & $2.64 \times 10^9$ & 45.4\% \\ \cline{2-6} 
 & IE + SP & $1.31 \times 10^9$ & 24.4\% & $2.00 \times 10^9$ & 29.4\% \\ \hline
\multirow{2}{*}{Batch Update} & SE & $1.20 \times 10^9$ & 58.3\% & $2.95 \times 10^9$ & 81.1\% \\ \cline{2-6} 
 & IE + SP & $1.18 \times 10^9$ & 58.2\% & $2.92 \times 10^9$ & 80.5\% \\ \hline
\end{tabular}

\end{table}

Additionally, we record the cache miss rates for software prefetching with interleaved execution and sequential execution. 
As shown in Table \ref{table:cache_miss}, the application of software prefetching reduced both cache misses and cache stall percentages. Specifically, in graph algorithms, cache misses were reduced by up to 35\%.

\section{Related Work}

\textbf{Dynamic graph storage systems.}
% \ys{
There have been graph storage systems developed for dynamic graph storage and processing ~\cite{pandey2021terrace,fuchs2022sortledton,DBLP:conf/icde/MackoMMS15,dhulipala2019low,de2021teseo,zhu2019livegraph,DBLP:journals/corr/abs-2004-00803,DBLP:journals/tos/KumarH20,DBLP:conf/hpec/WheatmanX18,kyrola2012graphchi}. 
% LiveGraph设计了日志结构类的结构来存储图数据，其为每个顶点分配一个独立边块连续的存储邻居以提供快速的邻居扫描，通过追加的方式来快速插入图更新。然而，这种追加的插入导致邻居无序的存储，无法高效的支持求交集等常用的算法需求。此外，当固定大小的边块数据装满时，其为了保证邻居的连续存储，需要将原始边块的数据复制到更大容量的比边块中，这时导致了插入性能的下降。
LiveGraph~\cite{zhu2019livegraph} adopts a log-structured design, allocating a contiguous edge block to each vertex for fast neighbor scanning and enabling swift updates through append-only insertions. However, this method leads to unordered neighbor storage, impeding the efficient execution of common operations, and data must be copied to larger blocks to maintain continuity once current edge blocks are full.\eat{, data must be copied to larger blocks to maintain continuity, reducing insertion speed.}
LLAMA~\cite{DBLP:conf/icde/MackoMMS15} uses multi-version array storage to enhance spatial locality in graph analysis. However, continuous data insertion leads to numerous array versions and scatters a vertex's neighbors across them, still causing extensive CPU cache stalls during analysis.
Aspen~\cite{dhulipala2019low} uses a tree structure for edge data to enable fast insertion and maintain ordered storage of neighbors. However, this structure results in minimal inherent spatial locality among data nodes, leading to poor access performance.
Terrace~\cite{pandey2021terrace} and Sortledton~\cite{fuchs2022sortledton} adopt different data structures for storing neighbors of vertices of varying degrees to enhance data locality and minimize cache misses during access. However, for vertices with a high degree, they are still stored in discontinuous structures, and neighbors of a single vertex may span multiple structures, ultimately impacting search and graph algorithm performance.
Teseo~\cite{de2021teseo} and PCSR~\cite{DBLP:conf/hpec/WheatmanX18}, employing CSR-like structures, use reserved gaps to prevent severe data movement overhead. However, they still face the same challenges as CSR when facing a surge in data updates.
Different from existing systems, we design a novel dynamic graph data structure to facilitate rapid graph insertions. It integrates the advantages of hardware prefetching, software prefetching, and coroutines to minimize cache misses, thereby enhancing graph computation performance.
% }

% LLAMA设计了多版本数组存储结构来存储图数据，为了在图分析时以利用数组良好的空间局部性。然而，随着数据的持续插入，会产生大量的版本的数组，并且一个顶点的邻居会分布在多个版本的数组中，这依然会导图分析时遭遇大量的CPU cache stall.

% Aspen等采用树状结构来存储边数据以支持快速的插入并保持邻居的有序存储，然而这些结构导致数据节点之间几乎没有固有的空间局部性，导致差的访问性能. 
% Terrace和Sortledton针对不同度数的顶点采用不同的数据结构来存储来提升数据的局部性以减少访问时的cache miss。然而，然而在顶点度数较大时，他们依然会存储在不连续的结构，并甚至可能一个顶点的邻居需要跨越多种结构存储,最终影响差找和图算法性能。
% Teseo、PCSR等采用类CSR结构，使用预留空隙来预防严重的data movements开销，但在更新数据量激增时，也无法避免与CSR相同的问题。
% 与现有的方法不同，我们设计新颖的动态图数据结构去支持快速的图插入。其充分结合硬件预取、软件预取和协程的特点，以减少动态图处理中的缓存丢失，从而提高图计算性能。

\textbf{Prefetching and Coroutines.}
% 由于CPU计算速度和内存访问延迟之间巨大差异，现有的一些系统例如\ref{terrace}采用预取来期望减少图分析过程中的CPU cache stall。然而这些同系统仅仅简单的利用系统提供的预取接口进行预取，并不能保证数据在访问前提取到cache中，因此依然遭受了严重的CPU cache stall开销。
% \ys{
Due to the significant disparity between CPU computation speeds and memory access latency, some existing graph systems~\cite{pandey2021terrace,fuchs2022sortledton,DBLP:conf/icde/MackoMMS15,dhulipala2019low,de2021teseo} employ prefetching with the aim of reducing CPU cache stalls during graph analysis.
However, these systems merely utilize the prefetching interfaces provided by the system in a straightforward manner, which does not guarantee that data is fetched into the cache prior to access. 
As a result, they still incur significant CPU cache stall overheads.
% 此外，一些工作采用基于协程的预取来加速现有的应用。例如CoroBase采用协程加速事物的处理过程。CoroGraph\ref{}中采用了coroutine-based prefetch来减少静态图系统上GAS过程中的CPU cache stall。相比于这些系统，我们针对动态图场景下的数据插入和图分析带来的独特挑战，结合Prefetch and Coroutines的特点设计了新颖的动态图结构以及访问模式，在支持快速的图插入的同时提供图分析的性能。
Furthermore, some works~\cite{he2020corobase,corograph} have adopted coroutine-based prefetching to accelerate existing applications. For example, CoroBase~\cite{he2020corobase} utilizes coroutines to expedite the transaction processing. In CoroGraph~\cite{corograph}, coroutine-based prefetching is employed to reduce CPU cache stalls during the Gather-Apply-Scatter (GAS) process in static graph systems.
Compared to these systems, our approach addresses the unique challenges brought by data insertion and graph analysis in dynamic graph scenarios. By integrating the characteristics of prefetching and coroutines, we design a novel dynamic graph structure and access pattern that supports rapid graph data insertion while enhancing graph analysis performance.
% }

\section{Conclusion}
\vspace{0.05in}
This paper presents \oursys{}, a prefetch-aware graph storage and software prefetching co-designed system by leveraging coroutines for dynamic graph processing.
C++20 coroutines are introduced to the system to achieve prefetching randomly stored graph data and reduce cache misses.
To benefit more from the software prefetching, several workload-balancing strategies and a prefetch-friendly data structure, \ourstorage, are proposed.
Extensive experiments on various datasets show the superiority of \oursys{} in achieving the balance of graph computation performance and graph update performance.

\begin{acks}
 % This work was supported by the [...] Research Fund of [...] (Number [...]). Additional funding was provided by [...] and [...]. We also thank [...] for contributing [...].
 This work was supported by the National Key R\&D Program of China (2023YFB4503601), the National Natural Science Foundation of China (U2241212, 62072082, and 62202088), the 111 Project (B16009), the Joint Funds of Natural Science Foundation of Liaoning Province (2023-MSBA-078), the Fundamental Research Funds for the Central Universities (N2416011), the Distinguished Youth Foundation of Liaoning Province (2024021148-JH3/501), and a research grant from Alibaba Innovative Research (AIR) Program. 
 % Hongfu Li and Qian Tao contributed equally. 
 Yanfeng Zhang and Shufeng Gong are the co-corresponding authors.
 % We thank the reviewers for their comments and suggestions.
 % We also thank our colleagues, friends, and family for their support and understanding.
 
 \eat{the National Key R\&D Program of China (2023YFB4503601), the National Natural Science Foundation of China (U2241212, 62072082, and 62202088), the 111 Project (B16009), the Fundamental Research Funds for the Central Universities (N2216015, N2216012), and a research grant from Alibaba Innovative Research (AIR) Program. Hongfu Li and Qian Tao contributed equally. Yanfeng Zhang and Shufeng Gong are the co-corresponding authors.}
 
% Natural Science Foundation of China (U2241212, 62072082, and 62202088), the 111 Project (B16009),  the Fundamental Research Funds for the Central Universities (N2216015, N2216012), and a research grant from Alibaba Innovative Research (AIR) Program. Yanfeng Zhang is the corresponding author.
\end{acks}
%\clearpage

\newpage
\balance
\bibliographystyle{acm}
\bibliography{GastCoCo}


\begin{thebibliography}{52}

%%% the Web address.

\ifx \showCODEN    \undefined \def \showCODEN     #1{\unskip}     \fi
\ifx \showDOI      \undefined \def \showDOI       #1{#1}\fi
\ifx \showISBNx    \undefined \def \showISBNx     #1{\unskip}     \fi
\ifx \showISBNxiii \undefined \def \showISBNxiii  #1{\unskip}     \fi
\ifx \showISSN     \undefined \def \showISSN      #1{\unskip}     \fi
\ifx \showLCCN     \undefined \def \showLCCN      #1{\unskip}     \fi
\ifx \shownote     \undefined \def \shownote      #1{#1}          \fi
\ifx \showarticletitle \undefined \def \showarticletitle #1{#1}   \fi
\ifx \showURL      \undefined \def \showURL       {\relax}        \fi
% invisible to TeX
\providecommand\bibfield[2]{#2}
\providecommand\bibinfo[2]{#2}
\providecommand\natexlab[1]{#1}
\providecommand\showeprint[2][]{arXiv:#2}

\bibitem[\protect\citeauthoryear{??}{Ali}{2023}]%
        {AlibabaDAU}
 \bibinfo{year}{2023}\natexlab{}.
\newblock \bibinfo{title}{Alibaba DAU}.
\newblock \bibinfo{howpublished}{\url{https://news.futunn.com/en/post/33640347?level=1&data_ticket=1709021520459060}}.
\newblock


\bibitem[\protect\citeauthoryear{??}{Tao}{2024}]%
        {Taobao}
 \bibinfo{year}{2024}\natexlab{}.
\newblock \bibinfo{title}{Taobao}.
\newblock \bibinfo{howpublished}{\url{https://www.taobao.com/}}.
\newblock


\bibitem[\protect\citeauthoryear{Angles, Antal, Averbuch, Boncz, Erling, Gubichev, Haprian, Kaufmann, Larriba{-}Pey, Mart{\'{\i}}nez{-}Bazan, Marton, Paradies, Pham, Prat{-}P{\'{e}}rez, Spasic, Steer, Sz{\'{a}}rnyas, and Waudby}{Angles et~al\mbox{.}}{2020}]%
        {DBLPdel}
\bibfield{author}{\bibinfo{person}{Renzo Angles}, \bibinfo{person}{J{\'{a}}nos~Benjamin Antal}, \bibinfo{person}{Alex Averbuch}, \bibinfo{person}{Peter~A. Boncz}, \bibinfo{person}{Orri Erling}, \bibinfo{person}{Andrey Gubichev}, \bibinfo{person}{Vlad Haprian}, \bibinfo{person}{Moritz Kaufmann}, \bibinfo{person}{Josep~Llu{\'{\i}}s Larriba{-}Pey}, \bibinfo{person}{Norbert Mart{\'{\i}}nez{-}Bazan}, \bibinfo{person}{J{\'{o}}zsef Marton}, \bibinfo{person}{Marcus Paradies}, \bibinfo{person}{Minh{-}Duc Pham}, \bibinfo{person}{Arnau Prat{-}P{\'{e}}rez}, \bibinfo{person}{Mirko Spasic}, \bibinfo{person}{Benjamin~A. Steer}, \bibinfo{person}{G{\'{a}}bor Sz{\'{a}}rnyas}, {and} \bibinfo{person}{Jack Waudby}.} \bibinfo{year}{2020}\natexlab{}.
\newblock \showarticletitle{The {LDBC} Social Network Benchmark}.
\newblock \bibinfo{journal}{\emph{CoRR}}  \bibinfo{volume}{abs/2001.02299} (\bibinfo{year}{2020}).
\newblock


\bibitem[\protect\citeauthoryear{Azadifar, Rostami, Berahmand, Moradi, and Oussalah}{Azadifar et~al\mbox{.}}{2022}]%
        {DBLP:journals/cbm/AzadifarRBMO22}
\bibfield{author}{\bibinfo{person}{Saeid Azadifar}, \bibinfo{person}{Mehrdad Rostami}, \bibinfo{person}{Kamal Berahmand}, \bibinfo{person}{Parham Moradi}, {and} \bibinfo{person}{Mourad Oussalah}.} \bibinfo{year}{2022}\natexlab{}.
\newblock \showarticletitle{Graph-based relevancy-redundancy gene selection method for cancer diagnosis}.
\newblock \bibinfo{journal}{\emph{Comput. Biol. Medicine}}  \bibinfo{volume}{147} (\bibinfo{year}{2022}), \bibinfo{pages}{105766}.
\newblock


\bibitem[\protect\citeauthoryear{Backstrom, Huttenlocher, Kleinberg, and Lan}{Backstrom et~al\mbox{.}}{2006}]%
        {Livejournal_dataset}
\bibfield{author}{\bibinfo{person}{Lars Backstrom}, \bibinfo{person}{Dan Huttenlocher}, \bibinfo{person}{Jon Kleinberg}, {and} \bibinfo{person}{Xiangyang Lan}.} \bibinfo{year}{2006}\natexlab{}.
\newblock \showarticletitle{Group formation in large social networks: membership, growth, and evolution}. In \bibinfo{booktitle}{\emph{Proceedings of the 12th ACM SIGKDD international conference on Knowledge discovery and data mining ({KDD})}}. \bibinfo{pages}{44--54}.
\newblock


\bibitem[\protect\citeauthoryear{Baer and Chen}{Baer and Chen}{1991}]%
        {DBLP:conf/sc/BaerC91}
\bibfield{author}{\bibinfo{person}{Jean{-}Loup Baer} {and} \bibinfo{person}{Tien{-}Fu Chen}.} \bibinfo{year}{1991}\natexlab{}.
\newblock \showarticletitle{An effective on-chip preloading scheme to reduce data access penalty}. In \bibinfo{booktitle}{\emph{Proceedings Supercomputing}}. \bibinfo{publisher}{{ACM}}, \bibinfo{pages}{176--186}.
\newblock


\bibitem[\protect\citeauthoryear{Beamer, Asanovic, and Patterson}{Beamer et~al\mbox{.}}{2015}]%
        {DBLP:conf/iiswc/BeamerAP15}
\bibfield{author}{\bibinfo{person}{Scott Beamer}, \bibinfo{person}{Krste Asanovic}, {and} \bibinfo{person}{David~A. Patterson}.} \bibinfo{year}{2015}\natexlab{}.
\newblock \showarticletitle{Locality Exists in Graph Processing: Workload Characterization on an Ivy Bridge Server}. In \bibinfo{booktitle}{\emph{2015 {IEEE} International Symposium on Workload Characterization({IISWC})}}. \bibinfo{publisher}{{IEEE} Computer Society}, \bibinfo{pages}{56--65}.
\newblock


\bibitem[\protect\citeauthoryear{Boldi and Vigna}{Boldi and Vigna}{2004}]%
        {hollywoodUK}
\bibfield{author}{\bibinfo{person}{Paolo Boldi} {and} \bibinfo{person}{Sebastiano Vigna}.} \bibinfo{year}{2004}\natexlab{}.
\newblock \showarticletitle{The {W}eb{G}raph Framework {I}: {C}ompression Techniques}. In \bibinfo{booktitle}{\emph{Proceedings of the Thirteenth International World Wide Web Conference (WWW)}}. \bibinfo{publisher}{ACM Press}, \bibinfo{pages}{595--601}.
\newblock


\bibitem[\protect\citeauthoryear{Callahan, Kennedy, and Porterfield}{Callahan et~al\mbox{.}}{1991}]%
        {callahan1991software}
\bibfield{author}{\bibinfo{person}{David Callahan}, \bibinfo{person}{Ken Kennedy}, {and} \bibinfo{person}{Allan Porterfield}.} \bibinfo{year}{1991}\natexlab{}.
\newblock \showarticletitle{Software prefetching}.
\newblock \bibinfo{journal}{\emph{ACM SIGARCH Computer Architecture News}} \bibinfo{volume}{19}, \bibinfo{number}{2} (\bibinfo{year}{1991}), \bibinfo{pages}{40--52}.
\newblock


\bibitem[\protect\citeauthoryear{Chakaravarthy, Checconi, Murali, Petrini, and Sabharwal}{Chakaravarthy et~al\mbox{.}}{2017}]%
        {DBLP:journals/tpds/ChakaravarthyCM17}
\bibfield{author}{\bibinfo{person}{Venkatesan~T. Chakaravarthy}, \bibinfo{person}{Fabio Checconi}, \bibinfo{person}{Prakash Murali}, \bibinfo{person}{Fabrizio Petrini}, {and} \bibinfo{person}{Yogish Sabharwal}.} \bibinfo{year}{2017}\natexlab{}.
\newblock \showarticletitle{Scalable Single Source Shortest Path Algorithms for Massively Parallel Systems}.
\newblock \bibinfo{journal}{\emph{{IEEE} Trans. Parallel Distributed Syst.}} \bibinfo{volume}{28}, \bibinfo{number}{7} (\bibinfo{year}{2017}), \bibinfo{pages}{2031--2045}.
\newblock


\bibitem[\protect\citeauthoryear{Chen, Ailamaki, Gibbons, and Mowry}{Chen et~al\mbox{.}}{2007}]%
        {chen2007improving}
\bibfield{author}{\bibinfo{person}{Shimin Chen}, \bibinfo{person}{Anastassia Ailamaki}, \bibinfo{person}{Phillip~B. Gibbons}, {and} \bibinfo{person}{Todd~C. Mowry}.} \bibinfo{year}{2007}\natexlab{}.
\newblock \showarticletitle{Improving hash join performance through prefetching}.
\newblock \bibinfo{journal}{\emph{{ACM} Trans. Database Syst.}} \bibinfo{volume}{32}, \bibinfo{number}{3} (\bibinfo{year}{2007}), \bibinfo{pages}{17}.
\newblock


\bibitem[\protect\citeauthoryear{Collins, Sair, Calder, and Tullsen}{Collins et~al\mbox{.}}{2002}]%
        {DBLP:conf/micro/CollinsSCT02}
\bibfield{author}{\bibinfo{person}{Jamison~D. Collins}, \bibinfo{person}{Suleyman Sair}, \bibinfo{person}{Brad Calder}, {and} \bibinfo{person}{Dean~M. Tullsen}.} \bibinfo{year}{2002}\natexlab{}.
\newblock \showarticletitle{Pointer cache assisted prefetching}. In \bibinfo{booktitle}{\emph{Proceedings of the 35th Annual International Symposium on Microarchitecture (MICRO)}}. \bibinfo{publisher}{{ACM/IEEE} Computer Society}, \bibinfo{pages}{62--73}.
\newblock


\bibitem[\protect\citeauthoryear{Comer}{Comer}{1979}]%
        {DBLP:journals/csur/Comer79}
\bibfield{author}{\bibinfo{person}{Douglas Comer}.} \bibinfo{year}{1979}\natexlab{}.
\newblock \showarticletitle{The Ubiquitous B-Tree}.
\newblock \bibinfo{journal}{\emph{{ACM} Comput. Surv.}} \bibinfo{volume}{11}, \bibinfo{number}{2} (\bibinfo{year}{1979}), \bibinfo{pages}{121--137}.
\newblock


\bibitem[\protect\citeauthoryear{Conway}{Conway}{1963}]%
        {conway1963design}
\bibfield{author}{\bibinfo{person}{Melvin~E. Conway}.} \bibinfo{year}{1963}\natexlab{}.
\newblock \showarticletitle{Design of a separable transition-diagram compiler}.
\newblock \bibinfo{journal}{\emph{Commun. {ACM}}} \bibinfo{volume}{6}, \bibinfo{number}{7} (\bibinfo{year}{1963}), \bibinfo{pages}{396--408}.
\newblock


\bibitem[\protect\citeauthoryear{de~Moura and Ierusalimschy}{de~Moura and Ierusalimschy}{2009}]%
        {moura2009revisiting}
\bibfield{author}{\bibinfo{person}{Ana~L{\'{u}}cia de Moura} {and} \bibinfo{person}{Roberto Ierusalimschy}.} \bibinfo{year}{2009}\natexlab{}.
\newblock \showarticletitle{Revisiting coroutines}.
\newblock \bibinfo{journal}{\emph{{ACM} Trans. Program. Lang. Syst.}} \bibinfo{volume}{31}, \bibinfo{number}{2} (\bibinfo{year}{2009}), \bibinfo{pages}{6:1--6:31}.
\newblock


\bibitem[\protect\citeauthoryear{Dhulipala, Blelloch, and Shun}{Dhulipala et~al\mbox{.}}{2019}]%
        {dhulipala2019low}
\bibfield{author}{\bibinfo{person}{Laxman Dhulipala}, \bibinfo{person}{Guy~E. Blelloch}, {and} \bibinfo{person}{Julian Shun}.} \bibinfo{year}{2019}\natexlab{}.
\newblock \showarticletitle{Low-latency graph streaming using compressed purely-functional trees}. In \bibinfo{booktitle}{\emph{Proceedings of the 40th {ACM} {SIGPLAN} Conference on Programming Language Design and Implementation (PLDI)}}. \bibinfo{publisher}{{ACM}}, \bibinfo{pages}{918--934}.
\newblock


\bibitem[\protect\citeauthoryear{Ding, Zeng, and Yu}{Ding et~al\mbox{.}}{2020}]%
        {BolinDingKaiZengWenyuanYu}
\bibfield{author}{\bibinfo{person}{Bolin Ding}, \bibinfo{person}{Kai Zeng}, {and} \bibinfo{person}{Wenyuan Yu}.} \bibinfo{year}{2020}\natexlab{}.
\newblock \showarticletitle{Alibaba Sponsor Talk at VLDB}.
\newblock


\bibitem[\protect\citeauthoryear{Ediger, McColl, Riedy, and Bader}{Ediger et~al\mbox{.}}{2012}]%
        {ediger2012stinger}
\bibfield{author}{\bibinfo{person}{David Ediger}, \bibinfo{person}{Robert McColl}, \bibinfo{person}{E.~Jason Riedy}, {and} \bibinfo{person}{David~A. Bader}.} \bibinfo{year}{2012}\natexlab{}.
\newblock \showarticletitle{{STINGER:} High performance data structure for streaming graphs}. In \bibinfo{booktitle}{\emph{{IEEE} Conference on High Performance Extreme Computing (HPEC)}}. \bibinfo{publisher}{{IEEE}}, \bibinfo{pages}{1--5}.
\newblock


\bibitem[\protect\citeauthoryear{Eisenman, Cherkasova, Magalhaes, Cai, Faraboschi, and Katti}{Eisenman et~al\mbox{.}}{2016}]%
        {DBLP:conf/wosp/EisenmanCMCFK16}
\bibfield{author}{\bibinfo{person}{Assaf Eisenman}, \bibinfo{person}{Ludmila Cherkasova}, \bibinfo{person}{Guilherme Magalhaes}, \bibinfo{person}{Qiong Cai}, \bibinfo{person}{Paolo Faraboschi}, {and} \bibinfo{person}{Sachin Katti}.} \bibinfo{year}{2016}\natexlab{}.
\newblock \showarticletitle{Parallel Graph Processing: Prejudice and State of the Art}. In \bibinfo{booktitle}{\emph{Proceedings of the 7th {ACM/SPEC} International Conference on Performance Engineering ({ICPE})}}. \bibinfo{publisher}{{ACM}}, \bibinfo{pages}{85--90}.
\newblock


\bibitem[\protect\citeauthoryear{Falsafi and Wenisch}{Falsafi and Wenisch}{2014}]%
        {DBLP:series/synthesis/2014Falsafi}
\bibfield{author}{\bibinfo{person}{Babak Falsafi} {and} \bibinfo{person}{Thomas~F. Wenisch}.} \bibinfo{year}{2014}\natexlab{}.
\newblock \bibinfo{booktitle}{\emph{A Primer on Hardware Prefetching}}.
\newblock \bibinfo{publisher}{Morgan {\&} Claypool Publishers}.
\newblock


\bibitem[\protect\citeauthoryear{Feng, Ma, Li, Zhu, Cai, Han, and Chen}{Feng et~al\mbox{.}}{2020}]%
        {DBLP:journals/corr/abs-2004-00803}
\bibfield{author}{\bibinfo{person}{Guanyu Feng}, \bibinfo{person}{Zixuan Ma}, \bibinfo{person}{Daixuan Li}, \bibinfo{person}{Xiaowei Zhu}, \bibinfo{person}{Yanzheng Cai}, \bibinfo{person}{Wentao Han}, {and} \bibinfo{person}{Wenguang Chen}.} \bibinfo{year}{2020}\natexlab{}.
\newblock \showarticletitle{RisGraph: {A} Real-Time Streaming System for Evolving Graphs}.
\newblock \bibinfo{journal}{\emph{CoRR}}  \bibinfo{volume}{abs/2004.00803} (\bibinfo{year}{2020}).
\newblock


\bibitem[\protect\citeauthoryear{Fuchs, Giceva, and Margan}{Fuchs et~al\mbox{.}}{2022}]%
        {fuchs2022sortledton}
\bibfield{author}{\bibinfo{person}{Per Fuchs}, \bibinfo{person}{Jana Giceva}, {and} \bibinfo{person}{Domagoj Margan}.} \bibinfo{year}{2022}\natexlab{}.
\newblock \showarticletitle{Sortledton: a universal, transactional graph data structure}.
\newblock \bibinfo{journal}{\emph{Proc. {VLDB} Endow.}} \bibinfo{volume}{15}, \bibinfo{number}{6} (\bibinfo{year}{2022}), \bibinfo{pages}{1173--1186}.
\newblock


\bibitem[\protect\citeauthoryear{Gupta, Satuluri, Grewal, Gurumurthy, Zhabiuk, Li, and Lin}{Gupta et~al\mbox{.}}{2014}]%
        {DBLP:journals/pvldb/GuptaSGGZLL14}
\bibfield{author}{\bibinfo{person}{Pankaj Gupta}, \bibinfo{person}{Venu Satuluri}, \bibinfo{person}{Ajeet Grewal}, \bibinfo{person}{Siva Gurumurthy}, \bibinfo{person}{Volodymyr Zhabiuk}, \bibinfo{person}{Quannan Li}, {and} \bibinfo{person}{Jimmy Lin}.} \bibinfo{year}{2014}\natexlab{}.
\newblock \showarticletitle{Real-Time Twitter Recommendation: Online Motif Detection in Large Dynamic Graphs}.
\newblock \bibinfo{journal}{\emph{Proc. VLDB Endow.}} \bibinfo{volume}{7}, \bibinfo{number}{13} (\bibinfo{year}{2014}), \bibinfo{pages}{1379--1380}.
\newblock


\bibitem[\protect\citeauthoryear{He, Hu, Lai, Li, Li, Li, Liu, Luo, Lyu, Meng, et~al\mbox{.}}{He et~al\mbox{.}}{2023}]%
        {he2023graphscope}
\bibfield{author}{\bibinfo{person}{Tao He}, \bibinfo{person}{Shuxian Hu}, \bibinfo{person}{Longbin Lai}, \bibinfo{person}{Dongze Li}, \bibinfo{person}{Neng Li}, \bibinfo{person}{Xue Li}, \bibinfo{person}{Lexiao Liu}, \bibinfo{person}{Xiaojian Luo}, \bibinfo{person}{Binqing Lyu}, \bibinfo{person}{Ke Meng}, {et~al\mbox{.}}} \bibinfo{year}{2023}\natexlab{}.
\newblock \showarticletitle{GraphScope Flex: LEGO-like Graph Computing Stack}.
\newblock \bibinfo{journal}{\emph{arXiv preprint arXiv:2312.12107}} (\bibinfo{year}{2023}).
\newblock


\bibitem[\protect\citeauthoryear{He, Lu, and Wang}{He et~al\mbox{.}}{2020}]%
        {he2020corobase}
\bibfield{author}{\bibinfo{person}{Yongjun He}, \bibinfo{person}{Jiacheng Lu}, {and} \bibinfo{person}{Tianzheng Wang}.} \bibinfo{year}{2020}\natexlab{}.
\newblock \showarticletitle{CoroBase: Coroutine-Oriented Main-Memory Database Engine}.
\newblock \bibinfo{journal}{\emph{Proc. {VLDB} Endow.}} \bibinfo{volume}{14}, \bibinfo{number}{3} (\bibinfo{year}{2020}), \bibinfo{pages}{431--444}.
\newblock


\bibitem[\protect\citeauthoryear{{Intel Corporation}}{{Intel Corporation}}{2016}]%
        {Intel2016}
\bibfield{author}{\bibinfo{person}{{Intel Corporation}}.} \bibinfo{year}{2016}\natexlab{}.
\newblock \bibinfo{booktitle}{\emph{Intel 64 and IA-32 Architectures Software Developer Manuals}}.
\newblock
\newblock
\shownote{(Oct. 2016).}


\bibitem[\protect\citeauthoryear{ISO/IEC}{ISO/IEC}{2017}]%
        {C++20}
\bibfield{author}{\bibinfo{person}{ISO/IEC}.} \bibinfo{year}{2017}\natexlab{}.
\newblock \bibinfo{title}{Technical Specification — C++ Extensions for Coroutines.}
\newblock \bibinfo{howpublished}{\url{https: //www.iso.org/standard/73008.html}}.
\newblock


\bibitem[\protect\citeauthoryear{Jonathan, Minhas, Hunter, Levandoski, and Nishanov}{Jonathan et~al\mbox{.}}{2018}]%
        {jonathan2018exploiting}
\bibfield{author}{\bibinfo{person}{Christopher Jonathan}, \bibinfo{person}{Umar~Farooq Minhas}, \bibinfo{person}{James Hunter}, \bibinfo{person}{Justin~J. Levandoski}, {and} \bibinfo{person}{Gor~V. Nishanov}.} \bibinfo{year}{2018}\natexlab{}.
\newblock \showarticletitle{Exploiting Coroutines to Attack the "Killer Nanoseconds"}.
\newblock \bibinfo{journal}{\emph{Proc. {VLDB} Endow.}} \bibinfo{volume}{11}, \bibinfo{number}{11} (\bibinfo{year}{2018}), \bibinfo{pages}{1702--1714}.
\newblock


\bibitem[\protect\citeauthoryear{Ko{\c{c}}berber, Falsafi, and Grot}{Ko{\c{c}}berber et~al\mbox{.}}{2015}]%
        {kocberber2015asynchronous}
\bibfield{author}{\bibinfo{person}{Yusuf~Onur Ko{\c{c}}berber}, \bibinfo{person}{Babak Falsafi}, {and} \bibinfo{person}{Boris Grot}.} \bibinfo{year}{2015}\natexlab{}.
\newblock \showarticletitle{Asynchronous Memory Access Chaining}.
\newblock \bibinfo{journal}{\emph{Proc. {VLDB} Endow.}} \bibinfo{volume}{9}, \bibinfo{number}{4} (\bibinfo{year}{2015}), \bibinfo{pages}{252--263}.
\newblock


\bibitem[\protect\citeauthoryear{Kumar and Huang}{Kumar and Huang}{2020}]%
        {DBLP:journals/tos/KumarH20}
\bibfield{author}{\bibinfo{person}{Pradeep Kumar} {and} \bibinfo{person}{H.~Howie Huang}.} \bibinfo{year}{2020}\natexlab{}.
\newblock \showarticletitle{GraphOne: {A} Data Store for Real-time Analytics on Evolving Graphs}.
\newblock \bibinfo{journal}{\emph{{ACM} Trans. Storage}} \bibinfo{volume}{15}, \bibinfo{number}{4} (\bibinfo{year}{2020}), \bibinfo{pages}{29:1--29:40}.
\newblock


\bibitem[\protect\citeauthoryear{Kyrola, Blelloch, and Guestrin}{Kyrola et~al\mbox{.}}{2012}]%
        {kyrola2012graphchi}
\bibfield{author}{\bibinfo{person}{Aapo Kyrola}, \bibinfo{person}{Guy~E. Blelloch}, {and} \bibinfo{person}{Carlos Guestrin}.} \bibinfo{year}{2012}\natexlab{}.
\newblock \showarticletitle{GraphChi: Large-Scale Graph Computation on Just a {PC}}. In \bibinfo{booktitle}{\emph{10th {USENIX} Symposium on Operating Systems Design and Implementation (OSDI)}}. \bibinfo{publisher}{{USENIX} Association}, \bibinfo{pages}{31--46}.
\newblock


\bibitem[\protect\citeauthoryear{Leo and Boncz}{Leo and Boncz}{2021}]%
        {de2021teseo}
\bibfield{author}{\bibinfo{person}{Dean~De Leo} {and} \bibinfo{person}{Peter~A. Boncz}.} \bibinfo{year}{2021}\natexlab{}.
\newblock \showarticletitle{Teseo and the Analysis of Structural Dynamic Graphs}.
\newblock \bibinfo{journal}{\emph{Proc. {VLDB} Endow.}} \bibinfo{volume}{14}, \bibinfo{number}{6} (\bibinfo{year}{2021}), \bibinfo{pages}{1053--1066}.
\newblock


\bibitem[\protect\citeauthoryear{Luk and Mowry}{Luk and Mowry}{1996}]%
        {DBLP:conf/asplos/LukM96}
\bibfield{author}{\bibinfo{person}{Chi{-}Keung Luk} {and} \bibinfo{person}{Todd~C. Mowry}.} \bibinfo{year}{1996}\natexlab{}.
\newblock \showarticletitle{Compiler-Based Prefetching for Recursive Data Structures}. In \bibinfo{booktitle}{\emph{Proceedings of the Seventh International Conference on Architectural Support for Programming Languages and Operating Systems (ASPLOS)}}. \bibinfo{publisher}{{ACM} Press}, \bibinfo{pages}{222--233}.
\newblock


\bibitem[\protect\citeauthoryear{Macko, Marathe, Margo, and Seltzer}{Macko et~al\mbox{.}}{2015}]%
        {DBLP:conf/icde/MackoMMS15}
\bibfield{author}{\bibinfo{person}{Peter Macko}, \bibinfo{person}{Virendra~J. Marathe}, \bibinfo{person}{Daniel~W. Margo}, {and} \bibinfo{person}{Margo~I. Seltzer}.} \bibinfo{year}{2015}\natexlab{}.
\newblock \showarticletitle{{LLAMA:} Efficient graph analytics using Large Multiversioned Arrays}. In \bibinfo{booktitle}{\emph{31st {IEEE} International Conference on Data Engineering ({ICDE})}}. \bibinfo{publisher}{{IEEE} Computer Society}, \bibinfo{pages}{363--374}.
\newblock


\bibitem[\protect\citeauthoryear{Page, Brin, and Motwani}{Page et~al\mbox{.}}{1998}]%
        {page1998winograd}
\bibfield{author}{\bibinfo{person}{Larry Page}, \bibinfo{person}{Sergey Brin}, {and} \bibinfo{person}{R Motwani}.} \bibinfo{year}{1998}\natexlab{}.
\newblock \showarticletitle{Winograd., T. The pagerank citation ranking: bringing order to the web}.
\newblock \bibinfo{journal}{\emph{Unpublished manuscript}} (\bibinfo{year}{1998}).
\newblock


\bibitem[\protect\citeauthoryear{Pandey, Wheatman, Xu, and Bulu{\c{c}}}{Pandey et~al\mbox{.}}{2021}]%
        {pandey2021terrace}
\bibfield{author}{\bibinfo{person}{Prashant Pandey}, \bibinfo{person}{Brian Wheatman}, \bibinfo{person}{Helen Xu}, {and} \bibinfo{person}{Aydin Bulu{\c{c}}}.} \bibinfo{year}{2021}\natexlab{}.
\newblock \showarticletitle{Terrace: {A} Hierarchical Graph Container for Skewed Dynamic Graphs}. In \bibinfo{booktitle}{\emph{International Conference on Management of Data (SIGMOD)}}. \bibinfo{publisher}{{ACM}}, \bibinfo{pages}{1372--1385}.
\newblock


\bibitem[\protect\citeauthoryear{Psaropoulos, Legler, May, and Ailamaki}{Psaropoulos et~al\mbox{.}}{2017}]%
        {DBLP:journals/pvldb/PsaropoulosLMA17}
\bibfield{author}{\bibinfo{person}{Georgios Psaropoulos}, \bibinfo{person}{Thomas Legler}, \bibinfo{person}{Norman May}, {and} \bibinfo{person}{Anastasia Ailamaki}.} \bibinfo{year}{2017}\natexlab{}.
\newblock \showarticletitle{Interleaving with Coroutines: {A} Practical Approach for Robust Index Joins}.
\newblock \bibinfo{journal}{\emph{Proc. {VLDB} Endow.}} \bibinfo{volume}{11}, \bibinfo{number}{2} (\bibinfo{year}{2017}), \bibinfo{pages}{230--242}.
\newblock


\bibitem[\protect\citeauthoryear{Qiu, Cen, Qian, Peng, Zhang, Lin, and Zhou}{Qiu et~al\mbox{.}}{2018}]%
        {DBLP:journals/pvldb/QiuCQPZLZ18}
\bibfield{author}{\bibinfo{person}{Xiafei Qiu}, \bibinfo{person}{Wubin Cen}, \bibinfo{person}{Zhengping Qian}, \bibinfo{person}{You Peng}, \bibinfo{person}{Ying Zhang}, \bibinfo{person}{Xuemin Lin}, {and} \bibinfo{person}{Jingren Zhou}.} \bibinfo{year}{2018}\natexlab{}.
\newblock \showarticletitle{Real-time Constrained Cycle Detection in Large Dynamic Graphs}.
\newblock \bibinfo{journal}{\emph{Proc. VLDB Endow.}} \bibinfo{volume}{11}, \bibinfo{number}{12} (\bibinfo{year}{2018}), \bibinfo{pages}{1876--1888}.
\newblock


\bibitem[\protect\citeauthoryear{Roth, Moshovos, and Sohi}{Roth et~al\mbox{.}}{1998}]%
        {Roth_Moshovos_Sohi_1998}
\bibfield{author}{\bibinfo{person}{Amir Roth}, \bibinfo{person}{Andreas Moshovos}, {and} \bibinfo{person}{Gurindar~S. Sohi}.} \bibinfo{year}{1998}\natexlab{}.
\newblock \showarticletitle{Dependence based prefetching for linked data structures}. In \bibinfo{booktitle}{\emph{Proceedings of the Eighth International Conference on Architectural Support for Programming Languages and Operating Systems (ASPLOS)}}. \bibinfo{publisher}{{ACM} Press}, \bibinfo{pages}{115--126}.
\newblock


\bibitem[\protect\citeauthoryear{Roth and Sohi}{Roth and Sohi}{1999}]%
        {DBLP:conf/isca/RothS99}
\bibfield{author}{\bibinfo{person}{Amir Roth} {and} \bibinfo{person}{Gurindar~S. Sohi}.} \bibinfo{year}{1999}\natexlab{}.
\newblock \showarticletitle{Effective Jump-Pointer Prefetching for Linked Data Structures}. In \bibinfo{booktitle}{\emph{Proceedings of the 26th Annual International Symposium on Computer Architecture ({ISCA})}}. \bibinfo{publisher}{{IEEE} Computer Society}, \bibinfo{pages}{111--121}.
\newblock


\bibitem[\protect\citeauthoryear{Saad}{Saad}{2003}]%
        {DBLP:books/daglib/0009092}
\bibfield{author}{\bibinfo{person}{Yousef Saad}.} \bibinfo{year}{2003}\natexlab{}.
\newblock \bibinfo{booktitle}{\emph{Iterative methods for sparse linear systems}}.
\newblock \bibinfo{publisher}{{SIAM}}.
\newblock


\bibitem[\protect\citeauthoryear{Sahu, Mhedhbi, Salihoglu, Lin, and {\"{O}}zsu}{Sahu et~al\mbox{.}}{2017}]%
        {DBLP:journals/pvldb/SahuMSLO17}
\bibfield{author}{\bibinfo{person}{Siddhartha Sahu}, \bibinfo{person}{Amine Mhedhbi}, \bibinfo{person}{Semih Salihoglu}, \bibinfo{person}{Jimmy Lin}, {and} \bibinfo{person}{M.~Tamer {\"{O}}zsu}.} \bibinfo{year}{2017}\natexlab{}.
\newblock \showarticletitle{The Ubiquity of Large Graphs and Surprising Challenges of Graph Processing}.
\newblock \bibinfo{journal}{\emph{Proc. VLDB Endow.}} \bibinfo{volume}{11}, \bibinfo{number}{4} (\bibinfo{year}{2017}), \bibinfo{pages}{420--431}.
\newblock


\bibitem[\protect\citeauthoryear{Sharma, Jiang, Bommannavar, Larson, and Lin}{Sharma et~al\mbox{.}}{2016}]%
        {DBLP:journals/pvldb/SharmaJBLL16}
\bibfield{author}{\bibinfo{person}{Aneesh Sharma}, \bibinfo{person}{Jerry Jiang}, \bibinfo{person}{Praveen Bommannavar}, \bibinfo{person}{Brian Larson}, {and} \bibinfo{person}{Jimmy Lin}.} \bibinfo{year}{2016}\natexlab{}.
\newblock \showarticletitle{GraphJet: Real-Time Content Recommendations at Twitter}.
\newblock \bibinfo{journal}{\emph{Proc. {VLDB} Endow.}} \bibinfo{volume}{9}, \bibinfo{number}{13} (\bibinfo{year}{2016}), \bibinfo{pages}{1281--1292}.
\newblock


\bibitem[\protect\citeauthoryear{Smith}{Smith}{1978}]%
        {DBLP:journals/computer/Smith78}
\bibfield{author}{\bibinfo{person}{Alan~Jay Smith}.} \bibinfo{year}{1978}\natexlab{}.
\newblock \showarticletitle{Sequential Program Prefetching in Memory Hierarchies}.
\newblock \bibinfo{journal}{\emph{Computer}} \bibinfo{volume}{11}, \bibinfo{number}{12} (\bibinfo{year}{1978}), \bibinfo{pages}{7--21}.
\newblock


\bibitem[\protect\citeauthoryear{Tu, Qu, Wu, Zhang, Liu, Zhao, Wu, Zhou, and Zhang}{Tu et~al\mbox{.}}{2023}]%
        {DBLP:conf/cikm/TuQWZLZWZZ23}
\bibfield{author}{\bibinfo{person}{Ke Tu}, \bibinfo{person}{Wei Qu}, \bibinfo{person}{Zhengwei Wu}, \bibinfo{person}{Zhiqiang Zhang}, \bibinfo{person}{Zhongyi Liu}, \bibinfo{person}{Yiming Zhao}, \bibinfo{person}{Le Wu}, \bibinfo{person}{Jun Zhou}, {and} \bibinfo{person}{Guannan Zhang}.} \bibinfo{year}{2023}\natexlab{}.
\newblock \showarticletitle{Disentangled Interest importance aware Knowledge Graph Neural Network for Fund Recommendation}. In \bibinfo{booktitle}{\emph{Proceedings of the 32nd {ACM} International Conference on Information and Knowledge Management ({CIKM})}}. \bibinfo{publisher}{{ACM}}, \bibinfo{pages}{2482--2491}.
\newblock


\bibitem[\protect\citeauthoryear{Wang, Shen, Huang, Wu, Dong, and Kanakia}{Wang et~al\mbox{.}}{2020}]%
        {ogbl}
\bibfield{author}{\bibinfo{person}{Kuansan Wang}, \bibinfo{person}{Zhihong Shen}, \bibinfo{person}{Chiyuan Huang}, \bibinfo{person}{Chieh{-}Han Wu}, \bibinfo{person}{Yuxiao Dong}, {and} \bibinfo{person}{Anshul Kanakia}.} \bibinfo{year}{2020}\natexlab{}.
\newblock \showarticletitle{Microsoft Academic Graph: When experts are not enough}.
\newblock \bibinfo{journal}{\emph{Quant. Sci. Stud.}} \bibinfo{volume}{1}, \bibinfo{number}{1} (\bibinfo{year}{2020}), \bibinfo{pages}{396--413}.
\newblock


\bibitem[\protect\citeauthoryear{Wheatman and Xu}{Wheatman and Xu}{2018}]%
        {DBLP:conf/hpec/WheatmanX18}
\bibfield{author}{\bibinfo{person}{Brian Wheatman} {and} \bibinfo{person}{Helen Xu}.} \bibinfo{year}{2018}\natexlab{}.
\newblock \showarticletitle{Packed Compressed Sparse Row: {A} Dynamic Graph Representation}. In \bibinfo{booktitle}{\emph{2018 {IEEE} High Performance Extreme Computing Conference ({HPEC})}}. \bibinfo{publisher}{{IEEE}}, \bibinfo{pages}{1--7}.
\newblock


\bibitem[\protect\citeauthoryear{Yang and Leskovec}{Yang and Leskovec}{2012}]%
        {OrkutComFriend}
\bibfield{author}{\bibinfo{person}{Jaewon Yang} {and} \bibinfo{person}{Jure Leskovec}.} \bibinfo{year}{2012}\natexlab{}.
\newblock \showarticletitle{Defining and Evaluating Network Communities Based on Ground-Truth}. In \bibinfo{booktitle}{\emph{12th {IEEE} International Conference on Data Mining ({ICDM})}}. \bibinfo{publisher}{{IEEE} Computer Society}, \bibinfo{pages}{745--754}.
\newblock


\bibitem[\protect\citeauthoryear{Zamini, Reza, and Rabiei}{Zamini et~al\mbox{.}}{2022}]%
        {DBLP:journals/information/ZaminiRR22}
\bibfield{author}{\bibinfo{person}{Mohamad Zamini}, \bibinfo{person}{Hassan Reza}, {and} \bibinfo{person}{Minou Rabiei}.} \bibinfo{year}{2022}\natexlab{}.
\newblock \showarticletitle{A Review of Knowledge Graph Completion}.
\newblock \bibinfo{journal}{\emph{Inf.}} \bibinfo{volume}{13}, \bibinfo{number}{8} (\bibinfo{year}{2022}), \bibinfo{pages}{396}.
\newblock


\bibitem[\protect\citeauthoryear{Zhi, Yan, Tang, Yin, Zhu, and Zhou}{Zhi et~al\mbox{.}}{2024}]%
        {corograph}
\bibfield{author}{\bibinfo{person}{Xiangyu Zhi}, \bibinfo{person}{Xiao Yan}, \bibinfo{person}{Bo Tang}, \bibinfo{person}{Ziyao Yin}, \bibinfo{person}{Yanchao Zhu}, {and} \bibinfo{person}{Minqi Zhou}.} \bibinfo{year}{2024}\natexlab{}.
\newblock \showarticletitle{CoroGraph: Bridging Cache Efficiency and Work Efficiency for Graph Algorithm Execution}.
\newblock \bibinfo{journal}{\emph{Proc. {VLDB} Endow.}} \bibinfo{volume}{17}, \bibinfo{number}{4} (\bibinfo{year}{2024}), \bibinfo{pages}{891--903}.
\newblock


\bibitem[\protect\citeauthoryear{Zhu, Chen, Zheng, and Ma}{Zhu et~al\mbox{.}}{2016}]%
        {DBLP:conf/osdi/ZhuCZM16}
\bibfield{author}{\bibinfo{person}{Xiaowei Zhu}, \bibinfo{person}{Wenguang Chen}, \bibinfo{person}{Weimin Zheng}, {and} \bibinfo{person}{Xiaosong Ma}.} \bibinfo{year}{2016}\natexlab{}.
\newblock \showarticletitle{Gemini: {A} Computation-Centric Distributed Graph Processing System}. In \bibinfo{booktitle}{\emph{12th {USENIX} Symposium on Operating Systems Design and Implementation, ({OSDI})}}. \bibinfo{publisher}{{USENIX} Association}.
\newblock


\bibitem[\protect\citeauthoryear{Zhu, Serafini, Ma, Aboulnaga, Chen, and Feng}{Zhu et~al\mbox{.}}{2020}]%
        {zhu2019livegraph}
\bibfield{author}{\bibinfo{person}{Xiaowei Zhu}, \bibinfo{person}{Marco Serafini}, \bibinfo{person}{Xiaosong Ma}, \bibinfo{person}{Ashraf Aboulnaga}, \bibinfo{person}{Wenguang Chen}, {and} \bibinfo{person}{Guanyu Feng}.} \bibinfo{year}{2020}\natexlab{}.
\newblock \showarticletitle{LiveGraph: {A} Transactional Graph Storage System with Purely Sequential Adjacency List Scans}.
\newblock \bibinfo{journal}{\emph{Proc. {VLDB} Endow.}} \bibinfo{volume}{13}, \bibinfo{number}{7} (\bibinfo{year}{2020}), \bibinfo{pages}{1020--1034}.
\newblock


\end{thebibliography}

\end{document}